\documentclass[10pt, twocolumn]{revtex4}
\usepackage{graphicx}
\newcommand{\be}{\begin{equation}}
\newcommand{\bea}{\begin{eqnarray}}
\newcommand{\ee}{\end{equation}}
\newcommand{\eea}{\end{eqnarray}}
\newcommand{\eps}{\epsilon}

\usepackage{amsmath}
\begin{document}

\title{Effective Hamiltonians for rapidly driven many-body lattice systems}
\author{A.P. Itin$^{1,2}$,  M.I. Katsnelson$^{1,3}$}
\affiliation{ $^1$Radboud University Nijmegen, Institute for Molecules and Materials (IMM), The Netherlands, \\
$^2$Space Research Institute,  Russian Academy of Sciences, Moscow, Russia, \\
$^3$ Dept. of Theoretical Physics and Applied Mathematics, Ural Federal University, Ekaterinburg, Russia} 
\begin{abstract}
We consider 1D lattices described by  Hubbard or Bose-Hubbard models,  in the presence of  periodic high-frequency perturbations, such as uniform ac force or modulation of hopping coefficients. Effective Hamiltonians for interacting particles are derived using an averaging method resembling classical canonical perturbation theory. As is known, a high-frequency force may renormalize hopping coefficients, causing interesting phenomena  such as coherent destruction of tunnelling and creation of artificial gauge fields. We find explicitly  additional corrections to the effective Hamiltonians  due to interactions, corresponding to non-trivial processes such as single-particle density-dependent tunnelling, correlated pair hoppings, nearest neighbour interactions, etc. Some of these processes  arise also in multiband lattice models, and are capable to  give rise to a rich variety of quantum phases.  The apparent contradiction with other methods, e.g. Floquet-Magnus expansion, is explained. The results may be useful for designing effective Hamiltonian models in experiments with ultracold atoms, as well as in the field of ultrafast nonequilibrium magnetism. An example of manipulating exchange interaction in a Mott-Hubbard insulator is considered, where our corrections play an essential role.
\end{abstract}
\maketitle

The idea of engineering effective Hamiltonians using high-frequency perturbations probably goes back
to the famous Kapitza pendulum \cite{Bogolyubov,Kapitza, Stephenson, Stephenson2}.  In classical and celestial mechanics there are many examples of systems with separation of typical timescales on slow and fast ones,  and corresponding  perturbation methods were developed long time ago (see, e.g.,  \cite{AKN}). It is interesting to adopt these methods to the quantum realm, especially to lattice systems, where  high-frequency perturbations are often used, e.g. for construction of quantum simulators (well-controllable quantum systems for simulating  complicated condensed matter phenomena  \cite{Feynman, Sim,Lew}). 
\begin{figure}[h]
\includegraphics[width=72mm]{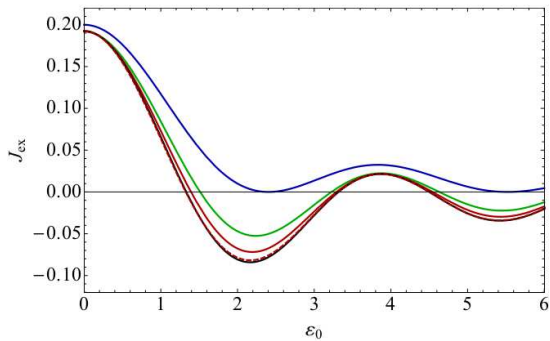}
\caption{ (Color online) Exchange interaction $J_{ex}$ in the driven fermionic two-site model as a function of the driving strength ${\cal E}_0$. Parameters: $U=10$, 'bare' tunnelling  $J=-1$, $\omega=16$.  Harmonic driving changes  effective tunnelling constant  $(J \to J_e =J J_0({\cal E}_0))$ affecting the exchange interaction in the zeroth order in $\frac{1}{\omega}$: $J_{ex}^{(0)}=\frac{2J_e^2({\cal E}_0)}{U}$. Solid curves,  from up to down: 'bare' exchange interaction $J_{ex}^{(0)}$; theoretical prediction $J_{ex}^{(2)}$ with corrections up to the second order in $\frac{1}{\omega}$  taken into account;  $J_{ex}^{(4)}$with corrections up to the fourth order in $\frac{1}{\omega}$  taken into account;   $J_{ex}^{(\infty)}$ with infinite order of terms in $\frac{1}{\omega}$ expansion taken into account  (with leading order in $U$ contribution in each term) (the lowest solid curve). Dashed curve (nearly inderscernable from the lowest solid curve):  numerical value of the exchange interaction (kindly provided by J.Mentink). In the regions  where  $J_e$  is strongly suppressed by driving (around zeros of $J_{ex}^{(0)}$ ), the second order corrections become insufficient, and one needs to include fourth and higher orders in $\frac{1}{\omega}$.
 \label{Fig1SI}}
\end{figure}
Indeed, a suitable driving applied to ultracold atoms in optical lattices  allows to  realize dynamical localization \cite{Arimondo1,Arimondo2}; mimick  photoconductivity \cite{Heinze}; simulate artificial gauge fields \cite{Struck, Gunea,Grechner}, classical and quantum magnetism \cite{Struck2}; study transport phenomena \cite{qtransport1, effective} and phase transitions  \cite{Eckardt, Zenesini2009}.
In solid-state physics,  ac-driven systems are interesting in the context of dynamical localization   \cite{Holthaus,Kenkre},  
coherent control of transport \cite{Korsch, Hohe};  microwave-induced topological insulators  \cite{Galitski};  photoinduced quantum Hall insulators   \cite{Kitagawa},  metal-insulator transitions \cite{Runi,Subidi}, and superconductivity \cite{Cavallery}.
The most of the abovementioned applications are in fact based on  modifying single-particle hopping amplitudes by a high-frequency force, a phenomenon that can be derived by averaging a Hamiltonian of the system (i.e., keeping only zeroth order terms in $\frac{1}{\omega}$ (inverse frequency of perturbation) and neglecting all higher-order terms. However, for many realistic applications of such type, it is important to derive accurate effective Hamiltonians taking into account higher-order terms (see, e.g., \cite{Longhi}). 
Here we determine explicitly  higher-order corrections to  effective Hamiltonians of driven  quantum lattices by elaborating a method \cite{AIN,Grozdanov,Rahav} inspired by canonical perturbation theory. The corrections correspond to non-trivial (many-body) processes such as single-particle density-dependent tunnelling, correlated pair hoppings,
nonlocal (extended) pair hopping, and so on.  Using a suitable driving, we are able to suppress or enhance a particular process in the effective Hamiltonian. Such approach can be very  useful for engineering particular Hamiltonians, for simulating solid-state phenomena via optical means, for accurate interpretation of experiments with driven lattice systems, etc.  A particular application to a recent  insightful proposal of ultrafast and reversible control of exchange interactions  \cite{Mentink} is demonstrated. 

The method can be seen either as a modification of the method of \cite{Rahav}, or as a modification of Magnus expansion approach \cite{Magnus,Blanes}. 
Surprisingly,  they can lead to different results. Correspondingly, some related methods available in the literature allow to obtain results that are in accord with ours \cite{Longhi, Rahav}, while others \cite{Blanes, Mielke} may lead to apparently different Hamiltonians. Especially the subtle difference between our approach and modifications of Magnus expansion method \cite{Blanes} is important. Magnus expansion \cite{Magnus,Blanes} is a popular tool in physics and mathematics with a rapidly growing number of applications.  It is interesting that sometimes accuracy of effective Hamiltonians produced by this method can be drastically increased, as explained below and in SI (note that Magnus expansion was actually designed for solving an initial value problem, not for deriving effective Hamiltonians). To clarify our approach, let us first start with  Schr\"odinger equation for a  lattice system  (e.g., a single particle in a  driven 1D tight-binding model), written in the matrix form:
\be   i \dot X = \eps {\cal H} X,   \label{H0} \ee  where 
$\eps = \frac{1}{\omega}$ is a small parameter, $\omega$ is a frequency of perturbation, ${\cal H}(t)$ is a time-dependent (matrix) Hamiltonian of the system, $X$ is a column of coefficients of expansion of a quantum state in a certain basis, and fast time was introduced ($t \to t/\eps $),  resulting in the small coefficient $\eps$ in front of r.h.s of Eq.(\ref{H0}), to put  high-frequency dependence of the Hamiltonian in explicit form. 
We may  consider Eq.(\ref{H0}) as a classical dynamical system for variables  $X_n$ (components of the column $X$). We adopt then a classical averaging method \cite{AKN, AIN, transport1,transport3,  IM} to this system,  in such a way that allows convenient generalization to many-body systems \cite{AIN}.
To this end, one makes a unitary  transformation  $X = C \tilde X$ so that equations for the transformed variables are
  \be i  \dot{ \tilde X} = [  C^{-1} \eps {\cal H} C - i C^{-1} \dot C  ]   \tilde X. \ee The expression in the square brackets is the new Hamiltonian. To get an effective time-independent Hamiltonian, the transformation is sought in the form  $ C = \exp[ \eps K_1 + \eps^2 K_2 + \eps^3 K_3+.. ], $ where $K_i$ are skew-Hermitian time-periodic matrices (with zero mean), which would remove time-dependent terms from the Hamiltonian, leaving only time-independent ones. 
An iterative procedure analogous to the Hamiltonian averaging method give us \cite{AIN}
 \bea  i  \dot{  K_1 }  &=&   {\cal H}(t)  -  \langle  {\cal H}(t) \rangle  \equiv  \left\{  {\cal H} \right\},   
  \quad i K_1  =  \int  \left\{  {\cal H} \right\} dt, \nonumber \\
  i  \dot{  K_2 }  &=& \left\{  {\cal H} K_1 - K_1 {\cal H} - \frac{i}{2} ( \dot{ K_1} K_1 - K_1 \dot{ K_1} )  \right\},  \nonumber  \\
\eps  {\cal H}_{eff} &=&  [  C^{-1} \eps  {\cal H} C - i C^{-1} \dot C  ]  =   \eps  {\cal H}_0 + \eps^2  {\cal H}_1 + .., \nonumber 
 \eea where curly brackets denote taking the time-periodic part of a time-dependent function : $\left\{ X \right\} \equiv  X -  \langle X(t) \rangle, $
where $\langle  X(t)   \rangle \equiv \frac{1}{2 \pi} \int \limits_0^{2\pi} X(t')  dt'$. Indefinite integrals above are defined up to an additive constant, which is chosen
in such a way that $\langle K_i \rangle =0$.
We have
 \bea
 {\cal H}_0 &=&   \langle  {\cal H}\rangle , \quad
 {\cal H}_1 =  \frac{1}{2} \langle [ \{  {\cal H} \} ,K_1]  \rangle,  \label{H123} \\
 {\cal H}_2 & =& \frac{1}{2} \langle [ \{ {\cal H} \}, K_2] \rangle    + \frac{1}{12}  \langle [ \{ [ \{ {\cal H} \},K_1] \},K_1 ] \rangle, .. \nonumber \eea 
% \bea
 %{\cal H}_0 &=&   \langle  {\cal H}\rangle , \quad
% {\cal H}_1 =  \frac{1}{2} \langle [ \{  {\cal H} \} ,K_1]  \rangle,  \label{H123} \\
% {\cal H}_2 & =& \langle [ {\cal H}, K_2] + \frac{1}{2}[ [ {\cal H},K_1],K_1 ]  - \frac{i}{2}( [\dot  K_1, K_2 ] \nonumber\\
% &+& [ \dot  K_2, K_1 ]  )  - \frac{i}{6} [ [ \dot{K_1},K_1],K_1]  \rangle, \nonumber
%\eea 
where square brackets denote matrix commutation: $[A,B] = AB -BA$.
This procedure resembles Floquet-Magnus expansion \cite{Blanes}, however there is an important difference due to  lifting unnecessary 
requirement $K_i(0)=0$  present in that method. As detailed in SI, it often allows to remove the correction ${\cal H}_1$, therefore obtaining much more accurate effective Hamiltonians, where corrections to ${\cal H}_0$ in the expansion start from ${\cal H}_2$.
 
Eqs.  (\ref{H123}) allow to consider, e.g., a particle in  driven tight-binding models with various boundary conditions, as well as with additional external potentials (see \cite{AIN} and SI).  The general  Eqs. (\ref{H123})  (which are in agreement with method of Ref. \cite{Grozdanov,Rahav} ) are also convenient for studying  many-body lattice systems.
Indeed,  in the case of many particles, one can construct a corresponding Hamiltonian matrix and fulfil  the same transformations.
Moreover, it is not necessary to consider Hamiltonians  (\ref{H123})  in the matrix representation: one can use, e.g., creation
and annihilation operators. Indeed, consider now 1D Bose-Hubbard model with a strong high-frequency driving:
\bea  H  &=&  H_{BH} + H_d(t),   \quad H_d(t)  =  \omega {\cal E} (\omega t) \sum \limits_j j n_j,   \nonumber\\
H_{BH}  &=&  J \sum \limits_i (c_i^{\dagger}  c_{i+1}+  c_{i+1}^{\dagger} c_i )    +  U \sum \limits_i n_i (n_i-1)    
\eea where $J$ is the hopping parameter, $U$ is the interaction strength, $c^{\dagger},c$ are bosonic creation and annihilation operators. 
Following the approach we used in the single-particle case, we make a preliminary transformation
 $U^{(0) }(t)=   \exp[i f(\omega t)    \sum \limits_j n_j  ] $, $f(\omega t) \equiv \int\limits^t_0 \omega {\cal E}(\omega t') dt' $,
 and make rescaling of time, $t' = \omega t = t/\eps$, so that the new Hamiltonian is
$ \eps  {\cal H}(t) = \eps ( {\cal H}_0  + \delta H(t) ), $
 where
\bea
{\cal H}_0 &=& \sum \limits_i J_{eff} ( c_i^{\dagger}  c_{i+1}+  c_{i+1}^{\dagger} c_i ) +  U \sum \limits_i n_i (n_i-1),  \nonumber\\
\delta H(t) & = &  \sum \limits_i  [ \delta^+(t) c_i^{\dagger}  c_{i+1}  +  \delta^- (t)   c_{i+1}^{\dagger} c_i  ], \\
J_{eff}  &=& J \langle e^{i f(t)} \rangle, \quad \delta^{\pm}(t) = J[ e^{\pm i f(t)} - \langle e^{i f(t)} \rangle]    \nonumber
\eea
In other words, the new Hamiltonian  is
$
\eps  {\cal H}(t) = \eps \Bigl[  \sum \limits_i  ( \delta_0^+(t) c_i^{\dagger}  c_{i+1}  +  \delta_0^- (t)   c_{i+1}^{\dagger} c_i   ) +  U \sum \limits_i n_i (n_i-1) \Bigr],
$
where $ \delta_0^{\pm} =  J e^{\pm i f(t)}.$

 The unitary transformations we fulfilled in the single-particle case should be done here as well.
 However, to find operators $K_1, K_2,  {\cal H}_1, {\cal H}_2$ explicitly, we fulfil more complicated many-body calculations.
 In the first order,  we have
 $  \dot{  K_1 }  =  -i   \left\{  {\cal H} \right\} = -i \eps  \sum \limits_i  [ \delta^+(t) c_i^{\dagger}  c_{i+1}  +  \delta^- (t)   c_{i+1}^{\dagger} c_i  ], $
 and therefore  $  K_1 = -i  \eps  \sum \limits_i  [ \delta_1^+(t) c_i^{\dagger}  c_{i+1}  +  \delta_1^- (t)   c_{i+1}^{\dagger} c_i  ] ,  $
 $\delta_1^{\pm} = \int \delta^{\pm} (t')dt'.$
 Commutators $[ {\cal H},K_1],  [[ {\cal H},K_1],K_1], $ etc,  are derived in SI.
The first-order terms in the new Hamiltonian (before averaging) are
 $
 [ {\cal H}, K_1] = -2 \eps iU \sum \limits_j  \delta_1^+   c_j^{\dagger}(n_j -n_{j+1}) c_{j+1}  -   \delta_1^-  c_{j+1}^{\dagger}(n_j-n_{j+1})c_j.$
 These terms resemble somehow the result of calculations of \cite{Mielke} (the structure of Hamiltonian is the same, but the 
 coefficients $\delta_1^{\pm}$ are different).  However, in our approach  this contribution disappears during time-averaging (being averaged to zero: ${\cal H}_1= \frac{1}{2} \langle [ \{  {\cal H} \} ,K_1]  \rangle =0$), and therefore to find the  non-vanishing contribution to the effective Hamiltonian, we need to consider the next orders of perturbation, exactly as in the single-particle case.  This  gives us (see SI)
 
\be {\cal H}_2 = -2U ( \Delta^+  \hat{a}_1  + h.c ) - 2U \Delta_0 \hat{a}_2,  \label{extended} \ee
 where  \bea
\Delta^+ &=&  \frac{1}{2}   \langle  \delta_2^+  \delta^+  \rangle, \quad
\Delta_0 =  \frac{1}{2} \langle    \delta^+ \delta_2^-  + \delta_2^+  \delta^-    \rangle,   \nonumber\\
\hat{a}_1  &=&   \sum \limits_j  \Bigl(   c_{j-1}^{\dagger}( 4n_j -n_{j+1} - n_{j-1}) c_{j+1}  \nonumber\\  &-&   2 c_j^{\dagger} c_j^{\dagger} c_{j+1} c_{j+1} \Bigr), \quad \delta_2^{\pm} = \int \delta_1^{\pm} (t')dt',  \\
\hat{a}_2 &=& \sum \limits_j  \Bigl[ 4n_j n_{j+1} - 2n_j (n_j-1) - (c_{j-1}^{\dagger} c_{j+1}^{\dagger} c_{j}^2  + H.c.) \Bigr] \nonumber
\eea

The effective Hamiltonian  contains  nearest-neighbour interactions and several types of correlated tunnelling processes (pair tunnelling $c_j^{\dagger} c_j^{\dagger} c_{j+1} c_{j+1}$ and pair "dissociation''/"association''  process  $c_{j-1}^{\dagger} c_{j+1}^{\dagger} c_{j} c_{j}$/ $c_j^{\dagger} c_j^{\dagger} c_{j+1}   c_{j-1} $).
Due to these terms, one may expect a rich phase diagram of the driven system. In particular,  it is known that extended Bose-Hubbard model,  obtained by adding 
nearest-neighbour interactions to the Bose-Hubbard Hamiltonian, poseses a supersolid phase.   

 We note that in the particular case of harmonic perturbation ${\cal E} = - {\cal E}_0 \sin t$, we have   $\Delta^+ = - \sum \limits_{k=1}^{\infty} (-1)^k \frac{J_k^2({\cal E}_0)}{k^2}$. This is a decaying oscillatory function of ${\cal E}_0$ which can be either positive or negative, and by varying the amplitude ${\cal E}_0$ one can either maximize its absolute value (e.g., at ${\cal E}_0=1.77$), or put it to zero (e.g., at ${\cal E}_0=3.33$). At the same time, $\Delta_0 = - \sum \limits_{k=1}^{\infty}  \frac{J_k^2({\cal E}_0)}{k^2} $, which is a non-vanishing oscillatory function of ${\cal E}_0$ with local minima at ${\cal E}_0=1.93, 5.32$, etc. We see that using a suitable driving,  it is possible to suppress or enhance particular processes in the effective Hamiltonian.
 
 %It is not difficult to show that using a more complicated driving with several parameters one can simultaneously put $\Delta_1$ and $\Delta_2$ to zero,
 %or selectively nulify one of them while  maximizing another one. This can be useful for making more precise experiments on {\em dynamical localization}.
 %In case of harmonic driving, a single-particle tunnelling is  stopped at $J_0(K)=0$.
 
It is worth to consider a classical limit $n_i \gg 1$, where one gets a driven Discrete Nonlinear Schr\"{o}dinger Equation (DNLSE). Applying canonical perturbation theory to the driven DNLSE we get an effective Hamiltonian which is indeed the classical limit of Eq.(\ref{extended}), i.e.
it can be obtained from the  Eq.(\ref{extended}) by replacing operators with $c-$numbers (SI). 
 
Consider now the driven Hubbard model \bea  H  &=&  H_{H} + H_d(t),  \quad H_d(t)  =  \omega {\cal E} (\omega t) \sum \limits_j j n_j,\\
H_{H}  &=&  J \sum \limits_{i,\sigma} (c_{i,\sigma}^{\dagger}  c_{i+1,\sigma}+  c_{i+1,\sigma}^{\dagger} c_{i,\sigma} )    +  U \sum \limits_i n_{i,\sigma} n_{i,-\sigma}  \nonumber   \eea (with $c_{i,\sigma}^{\dagger},c_{i,\sigma}^{\dagger}$ being fermionic creation and annihilation operators) which becomes,  after the preliminary transformation  discussed above, 
 \be
 {\cal H}  = \sum \limits_{i,\sigma} \Bigl(  \delta_0^+ c_{i,\sigma}^{\dagger} c_{i+1,\sigma}  + \delta_0^- c_{i+1,\sigma}^{\dagger} c_{i,\sigma}     \Bigr)  + U \sum  \limits_{i}  n_{i,\sigma} n_{i,-\sigma}. 
\nonumber \label{hubbard}
 \ee Calculations  analogous to the Bose-Hubbard model case give us (see SI) \bea
{\cal H}_1 &=&  0, \qquad {\cal H}_2 =  U \Delta_0 (-2{\cal S} + {\cal A} +{\cal A}^{\dagger} + 4 {\cal \bf V} -2 V )   \nonumber\\
 &+& U \Bigl[   \Delta^-  (4  {\cal \bf R} -2{\cal R}  +R_2 ) +H.c.Ê\Bigr], \label{FermiH}
\eea where   $ \Delta^-  =  (\Delta^+)^*,$ and several types of tunnelling and interaction processes were denoted
 as ${\cal S, A,R}, {\bf R}, R_2, V,$ and ${\cal \bf V}$ . ${\cal S}$ is the kinetic exchange interaction. ${\cal A}^{\dagger}$   denotes  a  tunnelling process where a pair of atoms at $j$th site is dissociated into two atoms at neighbouring sites ($j-1$ and $j+1$). ${\cal A}$ denotes a corresponding correlated tunnelling process of association of two atoms  into a pair. 
  ${\cal R}$ (${\cal R}^{\dagger}$)  denotes  a correlated tunnelling process where an {\em extended pair} of atoms is tunnelling to the right (left) neighbouring sites. $ {\cal \bf R}$ (${\cal \bf R}^{\dagger}$) denotes tunnelling of a localised pair of atoms to the right (left). $R_2$ ($R_2^{\dagger}$ ) denotes  single-particle next-nearest-neighbour density-dependent tunnelling to the right (left); $V$ is the nearest-neighbour interaction, and  $ {\cal \bf V}$ is  the usual local interaction.   
 \bea 
{\cal S} &=&   \sum_{j,\sigma} c_{j+1,\sigma}^{\dagger} c_{j,\sigma} c_{j,-\sigma}^{\dagger} c_{j+1,-\sigma},     \\
{\cal A}  &=& {\cal A}_{\uparrow, \downarrow}  +  {\cal A}_{\downarrow, \uparrow },   \quad  {\cal A}_{\sigma, -\sigma} = \sum_j c_{j,\sigma}^{\dagger} c_{j-1,\sigma} c_{j,-\sigma}^{\dagger} c_{j+1,-\sigma}, \nonumber\\
 {\cal R} &=&  {\cal R}_{  \uparrow \downarrow} +  {\cal R}_{ \downarrow \uparrow},  \quad {\cal R}_{ \sigma,- \sigma} = \sum_j c_{j,\sigma}^{\dagger} c_{j-1,\sigma} c_{j+1,-\sigma}^{\dagger} c_{j,-\sigma},  \nonumber\\
 {\cal \bf R} &=&  \sum_j c_{j+1,\downarrow}^{\dagger} c_{j,\downarrow} c_{j+1,\uparrow}^{\dagger} c_{j,\uparrow} ,  \nonumber\\
  R_2 &=&   \sum_{j,\sigma} c_{j+1,\sigma}^{\dagger} c_{j-1,\sigma}  (n_{j-1,-\sigma}- 2n_{j,-\sigma} +n_{j+1,-\sigma}),   \nonumber\\
V & =&  \sum_{j,\sigma} n_{j,\sigma} n_{j+1,-\sigma}, \quad {\cal \bf V} = \sum_j n_{j,\downarrow} n_{j,\uparrow}  \nonumber
 \eea
 After returning to the physical time (remember that we did rescaling $t \to t/\eps$),  the second-order correction is $\frac{1}{\omega^2} {\cal H}_2$ (Eq. \ref{FermiH}) (see also \cite{note}).
 
 All the second-order corrections we found enter the effective Hamiltonian with a prefactor $\frac{U J^2}{\omega^2}$. We therefore require
 not only  $\omega \gg J $, but also $\frac{U J^2}{\omega^2} \ll 1$ for our theory be applicable. 
It is clear that the larger $U$ makes the corrections more important. Moreover, it is possible to find situations where our "corrections" drastically change effective low-energy Hamiltonians.  E.g., for $U \gg 1$ and half-filling, the Hubbard model describes a Mott insulator where remaining spin degrees of freedom are coupled by an antiferromagnetic exchange interaction $J_{ex} = \frac{2J^2}{U}$
(this is captured already in a two-site Hubbard model: among the four states (for total $S_z=0$), the two low-lying states are singlet and triplet states with one electron per site and energies $E_S = - {\frac4 J^2}{U}$,  $E_T=0$. The spectrum at low energies is described by a spin Hamiltonian $2J_{ex} {\bf s}_1 {\bf s}_2$ with $J_{ex}= (E_T - E_S)/2 = \frac {2J^2}{U}$).   If parameters in the driven Hubbard model are such that  $\frac{J^2}{U} \sim U \frac{J^2}{\omega^2}$, i.e. $U \sim \omega$,  then high-frequency corrections are of the same order as the effective exchange interaction itself, and therefore play a crucial role. 
Very recently, it was proposed to control exchange interactions by time-periodic modulation of an electric field \cite{Mentink}.  The corresponding results of \cite{Mentink} can be reproduced and enlightened  in our approach.  For the two-site model, the correction to the effective Hamiltonian simplifies to

$A({\cal E}_0) \Bigr[  (n_{1\downarrow} - n_{2\downarrow})(n_{1\uparrow}-n_{2\uparrow}) - (c_{2\uparrow}^{\dagger}  c_{1\uparrow} c_{1\downarrow}^{\dagger} c_{2\downarrow} +  c_{2\downarrow}^{\dagger}  c_{1\downarrow} c_{1\uparrow}^{\dagger} c_{2\uparrow} )  \Bigl] -  C({\cal E}_0) ( d_2^{\dagger} d_1 + h.c.), $
where $d_k$ are annihilation operators of pairs (doublons), $A$ and $C$ are functions of driving that in the particular case of harmonic driving are equal to $-\frac{4U J^2}{\omega^2} \sum_1^{\infty} \frac{J_m^2({\cal E}_0)}{m^2} +O(\frac{1}{\omega^4})$ and $\frac{4U J^2}{\omega^2} \sum_1^{\infty} \frac{(-1)^m J_m^2({\cal E}_0)}{m^2}+O(\frac{1}{\omega^4})$, correspondingly.  Parameter $A$ (which gives the strength of exchange  interaction) can be calculated taking into account infinite number of terms in $\frac{1}{\omega}$ expansion,
if we leave only the leading order contribution in $U$ in every term. Then,  $A = -4J^2 U \sum \limits_{m=1}^{\infty} \frac{J_m^2({\cal E}_0)}{m^2 \omega^2-U^2}$.
It is interesting that this result, obtained under assumption $\omega >U$, remains valid also for $\omega <U$ \cite{Mentink}. We can obtain even more accurate expression
for the effective interaction,  which includes influence of other terms in the effective Hamiltonian:
eigenvalues of the effective two-site model can be obtained analytically (see SI). The half-distance between the two lowest levels is equal to $\Delta E/2 \equiv J_{ex} = 
\frac{1}{4} \left(  A+C-U + \sqrt{ 16J_e^2 + (U+3A -C)^2 }   \right) \approx  \frac{2J_e^2}{U} + A + \frac{2J_e^2}{U^2} (C-3A) + .. 
$. 
 For parameters used in \cite{Mentink}, $J_{ex}$ is shown on Fig. 1.  It  reproduces  calculations of  \cite{Mentink} amazingly well (see Fig.1 and SI).  Moreover, for longer lattices, our theory provides a value of the effective exchange interaction, which is free from (possible) finite-size effects of the two-site model. 
Indeed,  for large $U$, one can see that the most important contribution to the effective Hamiltonian comes from ${\cal S}$, other terms  increase the number of double occupancies and therefore are suppressed by the (original) interaction term of the Hubbard model.
 One can eliminate them via a standard approach, by a canonical transformation which renormalizes coefficients of the remaining terms, i.e. exchange interaction  (\cite{Girvin}).  When $\frac{J^2}{U} \sim U \frac{J^2}{\omega^2}$,  with a good accuracy we find the effective exchange interaction
 is composed of the usual $\frac{J^2}{U}$ part  and our correction from ${\cal S}$. So the driving not only modifies the  $\frac{J^2}{U}$ part  (by renormalizing $J \to J_{e}$), but also adds a 'correction' ${\cal S}$ which can be of the same order.  Suppression of the other terms in the effective Hamiltonian at large $U$ justify  usage of the two-site model result in the  extended system: we see that at least in the 1D case,  the change of the exchange interaction is just the same as in the two-site system. Recently, the fermonic two-site system  was realised experimentally \cite{Jochim}.  For a long $1D$ lattice,  the ability to switch exchange interaction to the ferromagnetic type imply
 an interesting possibility of simulating  an unusual  spin-polaron excitation in driven lattices:  a bound state of an extra fermion and a magnon \cite{Katsnelson}. Such a quasiparticle has peculiar properties, e.g. large effective mass (\cite{Katsnelson},SI).  We can also switch on and off  single-particle and doublon hopping by varying parameters of driving (SI).
  
%     insight not only to modification of the exchange interaction, but to other types of (driving-induced) effective interactions as well.  Note that at large $U$, the %most important contribution to the effective Hamiltonian comes from ${\cal S}$.
   
To conclude, a convenient method based on canonical (unitary) transformations has been applied to two different lattice  systems: driven 1D Hubbard and Bose-Hubbard models.   For a general high-frequency driving, we derive explicitly effective  Hamiltonians  including corrections from interactions.  For a Hubbard model,
a very interesting regime is found, where the corrections drastically influence the effective (Heisenberg) Hamiltonian.  

The results should be useful for forthcoming experiments with cold atoms in driven optical lattices.
In particular,  presently in experiments with shaken optical lattices utilising effective (averaged) tunnelling constants, a frequency of driving is typically chosen in kHz  regime, in order to be much higher than all  timescales related to trapping potential and interactions,  and much lower than  interband transition frequency.  Explicit knowledge of corrections from trapping potentials \cite{AIN} and many-body interactions we derived here allow to extend area of applicability of shaken lattice simulations to sub-kHz regime, by choosing driving protocols that cancel second-order corrections.  On the other hand,  one can amplify particular contributions from second-order corrections, thereby engineering new model Hamiltonians.  Another inspiring direction of applications  is  photoinduced superconductivity and metal-insulator transitions \cite{Runi,Subidi,Cavallery}, where recent  studies show importance of  induced Kapitza-like effective potentials \cite{Subidi}.  

Financial support by NWO via Spinoza Prize is acknowledged.
A.P.I is grateful to  A.I.Neishtadt, M. Eckstein, J.Mentink,  A.Polkovnikov,  S.Jochim,  A.Lichtenshtein, A.Nocera, J.Simonet,  M.Thorwart, T.Rexin  for inspiring discussions, and thanks K.Sengstock for hospitality during his stay in Hamburg. Partial support from RFBR (project no. 13-01-00251)  and LEXI (Hamburg) is acknowledged.

\newpage

\onecolumngrid
\section*{Supplementary Information}

\subsection{Comparison with Floquet-Magnus expansion}

Consider the linear differential equation
\be
Y'(t) = A(t) Y(t), \label{dY}
\ee 
where $A(t)$ is, for example,  a $n \times n$ matrix and $Y(t)$ is a n-dimensional vector function (more general objects can also be treated \cite{BlanesSI},
e.g. A(t) and Y(t) being operators in Hilbert space). Initial condition $ \quad Y(0)= Y_0=I  $ is imposed.
Magnus' proposal is to search for a solution of this initial value problem in the form of

\be
Y(t) =  \exp \left( \Omega(t) \right) Y_0 ,
\ee
where  $\Omega(t)$ is $n \times n$ matrix.

In the context of quantum physics, this approach is in contrast with the representation
$Y(t) =  {\cal T} \left( \exp \int \limits_0^t A(s) ds \right) $, where $ {\cal T}$ is the Dyson time-ordering operator.

Magnus found that
\be
\frac{d\Omega(t)}{dt} = \sum \limits_0^\infty \frac{B_n}{n!} ad_{\Omega}^n A,  \label{dom}
\ee
 where $ad_{\Omega}^n A$ is a shorthand for a nested commutator ($ad_{\Omega} A = [\Omega, A], \quad ad_{\Omega}^k A = [\Omega, ad_{\Omega}^{k-1} A]$),  and $B_n$ are the Bernoulli numbers.

Integration of (\ref{dom}) by means of iteration leads to a series for $\Omega$.
 In the original formulation,  this approach does not immediately  serve for finding time-independent Hamiltonians.
However, one can modify it in the following way \cite{BlanesSI}.
In the case of  periodically time-dependent matrix $A(t)$, according to Floquet and Lyapunov, a solution can be factorized on periodic part 
and a purely exponential factor:
\be
Y(t) = P(t) \exp(tF),    \label{factor}
\ee

where $F$ and $P(t)$ are $n \times n$ matrices, the latter is periodic $P(t+T)=P(t)$, and $F$ is constant.
Now,  one can interpretate  Eq.\ref{factor}  as following:  after transformation of the solution by  $ P^{-1}(t)$,  one get in the new representation
an equation of motion determined by constant matrix $F$ instead of $A(t)$.  Therefore,  $P(t)$ is the analog of our $C = \exp(\eps K_1(t) +..)$ from the main text,
while $F$ is the analog of our time-independent Hamiltonian $H_{eff}$.

Naively, $F$ and $H_{eff}$ should merely coincide with each other.  However, there is a subtle but important difference stemming from different initial conditions.

Let us  remind the derivation of Floquet-Magnus expansion \cite{BlanesSI}.
 Introducing the factorized form \ref{factor} into the differential equation  $Y'=A(t)Y$, one gets equation of motion for $P(t)$.

\be
P'(t) = A(t) P(t) - P(t) F,  \quad P(0)=I.
\ee
For $P(t)$ one uses the exponential ansatz 
\be
P(t) = \exp(\Lambda(t)), \quad \Lambda(0)=0,
\ee
 and obtains
 
 \be
 \Lambda' = \sum\limits_{k=0}^{\infty} \frac{B_k}{k!} ad_{\Lambda}^k (A +(-1)^{k+1} F)
 \ee
 Then, one considers series expansion for $F$ and $\Lambda$ 
 \be
 \Lambda(t) = \sum \limits_{k=1}^{\infty} \Lambda_k(t), \quad F = \sum \limits_{k=1}^{\infty} F_k, 
 \ee
 with all $\Lambda_k(0)=0$.   $\Lambda_k$ is supposed to be of order $\eps^k$. Alternatively, one can replace $A$ by $\eps A$, obtaining powers of  $\eps$ explicitly
 in the expansion $\Lambda(t) = \sum  \limits_{k=1}^{\infty} \eps^k \Lambda_k(t)$. 
 
 Evaluating terms with the same order, one gets
 
 \bea
 \Lambda_1(t) &=& \int_0^t A(x) dx  - t F_1, \nonumber\\
 F_1 &=& \frac{1}{T} \int_0^T A(x) dx,  \nonumber\\
 \Lambda_2(t)  &=& \frac{1}{2} \int \limits_0^t [ A(x) + F_1, \Lambda_1] dx - tF_2,  \nonumber\\
 F_2 &=& \frac{1}{2T} \int \limits_0^T [A(x) + F_1, \Lambda_1(x)] dx 
 \eea

 One can see that while $F_1$ coincide with our $H_0$,  $F_2$ is different from $H_1$.
 The reason is that the mean value of $\Lambda_k$ is not zero. Instead,  a condition $\Lambda_k=0$ is imposed.
 In fact, this condition is unnecessary for our purposes (for derivation of effective Hamiltonians).
 Let us return back to the original Magnus expansion. In the equation  (\ref{dY}),  the initial condition $Y(0)=I$ was imposed.  
General solution of (\ref{dY}) with other initial conditions can be obtained by multiplying $Y(t)$   by a constant matrix.
 The chosen initial condition  transforms to $P(0)=I$  when considering the periodic system, and then to $\Lambda(0)=0$.
 However,  generally we can admit  transformations where $P(0) \ne I$,  since we actually interested not in
 constructing the fundamental matrix for the solution of the initial value problem, but in constructing effective Hamiltonians. 
 So, its not a problem if  $\bar X(0) \ne X(0)$ in the main text.  Our main priority is that transformed $\bar X$ evolves according to 
 time-independent effective Hamiltonian which captures averaged behaviour of the system on long times as good as possible.
  This freedom ($\bar X(0) \ne X(0)$) allows for tremendous  improvement in the construction of the effective Hamiltonian:
 namely, cancellation of the first-order corrections.  The  remaining second-order corrections cannot be removed by further linear transformations
 and therefore essential for  understanding averaged properties of the system.  One necessarily need to truncate expansion of the time-independent matrix $F$  at some order. Clearly, it is insufficient to truncate it on the first order corrections ($F_2$), since this part can be removed as described above.
 We see that while our expansion described in the main text is in accord with Magnus expansion,  it is slightly different due to initial conditions, and is actually more effective for revealing long-time features of the system:  it completely removes the lowest order correction, so that only higher-order  corrections remains.  
  
 %Indeed all solutions $Y(t)$  of Eq.(\ref{dY})  can be obtained from (\ref{factor}) by multiplying R.H.S. of  (\ref{factor})  by a constant matrix $P_0$.

\subsection{A single-particle tight-binding model}

Here and in the next two sections we consider particular models and calculate commutators for the older version of  Eqs. (3) of the main text,
that is for expressions available in \cite{AINSI}:

\bea
 {\cal H}_0 &=&   \langle  {\cal H}\rangle , \quad
 {\cal H}_1 =  \frac{1}{2} \langle [ \{  {\cal H} \} ,K_1]  \rangle,  \label{AIN123} \\
 {\cal H}_2 & =& \langle [ {\cal H}, K_2] + \frac{1}{2}[ [ {\cal H},K_1],K_1 ]  - \frac{i}{2}( [\dot  K_1, K_2 ] \nonumber\\
 &+& [ \dot  K_2, K_1 ]  )  - \frac{i}{6} [ [ \dot{K_1},K_1],K_1]  \rangle, \nonumber
\eea 

Despite being less compact,  they allow somehow deeper insight in the structure of various terms of the expansion.

In the case of a particle in a fastly driven tight-binding chain, one has (see \cite{AINSI})
$ H  =  J \sum ( | n \rangle \langle n+1|    + | n+1 \rangle \langle n | )  -  ed \omega E(\omega t) \sum \limits_n n | n \rangle \langle n |, $
where  $d$ is the intersite distance, $\omega E$ is the applied electric field (here, we explicitly place $\omega \equiv 1/\eps$ in the definition to emphasize  strong driving), 
$e$ is the charge of the particle,  $J$ is the tunnelling constant. 
The same model can be realized also with neutral particles (in the co-moving frame), by appropriate shaking of the lattice \cite{EckardtSI} .
Expanding a quantum state as $| \psi (t) \rangle = \sum c_n | n \rangle,$ we  get a system of equations \be i \dot c_n = J (c_{n+1} +  c_{n-1} )  - 
\omega {\cal E}(\omega t) n c_n. \ee
We make a transformation $c_n(t) = X_n(t) \exp \Bigl[ - i n \int\limits^t_0 \omega {\cal E}(\omega t') dt'  \Bigr] $, so that equations of motion are

\be i \dot X_n = J (X_{n+1} F(\omega t) +  X_{n-1} F^*(\omega t)),   \label{EOM} \ee
where $F(\omega t) = \exp [ - i \int\limits_0^t {\omega \cal E}(\omega t') dt'  ] \  =  F_0 + \sum F_l  \exp(-i l \omega t),$   $ {\cal E} = edE$.

Introducing fast time $t'  = \omega t  \equiv t/\eps $, we get,  in the matrix form, Eq.(1) of the main text, with  the Hamiltonian   ${\cal H } $ corresponding to Eq.(\ref{EOM}). 

In a more general  setting,  a particle in a driven tight-binding chain subject to an additional external potential $V(n)$,
we have
\be i \dot X_n = \eps[ J (X_{n+1} F(t) +  X_{n-1} F^*(t)) + V(n) X_n],   \label{EOMV} \ee
or
\be i \dot X = \eps H,   \label{EOMVF} \ee

 In the spirit of classical canonical perturbation theory, we are making a  unitary transformation $X = C \tilde X$ so that equations for the transformed variables are
 
 \be i  \dot{ \tilde X} = [  C^{-1} \eps H C - i C^{-1} \dot C  ]   \tilde X . \ee
 
 We are searching for a transformation of the form $C = \exp[ \eps K_1 + \eps^2 K_2 + \eps^3 K_3 ], $
 where  $K_i$ are skew-Hermitian time-periodic  matrices, which would remove time-dependent terms from the Hamiltonian, leaving only time-independent terms.
 
We have 
 
 \bea  C &\approx &I + \eps K_1 + \eps^2 \left( \frac{1}{2} K_1^2 + K_2 \right) + \eps^3 \left( \frac{1}{6} K_1^3 +\frac{1}{2}(K_1 K_2 + K_2 K_1) + K_3 \right) , \nonumber\\
       C^\dagger &\approx& I - \eps K_1  + \eps^2 \left( \frac{1}{2} K_1^2 - K_2 \right) + \eps^3 \left( -\frac{1}{6} K_1^3 +\frac{1}{2}(K_1 K_2 + K_2 K_1) - K_3 \right),  \eea
 where $I$ is the unity matrix.
 
 In the first order,  we have
 
 \be  i  \dot{  K_1 }  =  H(t)  -  \langle H(t) \rangle  \equiv  \left\{ H \right\}, \ee
 
 and therefore  $ i K_1 = \int  (H - \langle H \rangle ) dt  =  \int  \left\{ H \right\} dt . $
We introduce here curly brackets as taking time-periodic part of a time-dependent function: $\left\{ X \right\} \equiv  X -  \langle X(t) \rangle, $
where $\langle  X(t)   \rangle \equiv \frac{1}{2 \pi} \int \limits_0^{2\pi} X(t')  dt'$.

In the second order,
 
 \be  i  \dot{  K_2 }  = \left\{ H K_1 - K_1 H - \frac{i}{2} ( \dot{ K_1} K_1 - K_1 \dot{ K_1} )  \right\},    \ee
and the effective Hamiltonian is
\be  \eps H_{eff} = \eps {\cal H}_0 + \eps^2 {\cal H}_1 + \eps^3 {\cal H}_2 ,\ee
where

\bea
{\cal H}_0 &=&   \langle H \rangle \nonumber\\
{\cal H}_1 &=& \langle H K_1 - K_1 H - \frac{i}{2} (\dot  K_1 K_1 - K_1 \dot  K_1)  \rangle  \label{formula} \\
{\cal H}_2 & =& \langle H K_2 - K_2 H + \frac{1}{2} (H K_1^2 + K_1^2 H) - K_1 H K_1  - \frac{i}{2}(\dot  K_1 K_2 - K_1 \dot  K_2 + \dot  K_2 K_1 - K_2 \dot  K_1  ) \nonumber\\
  &-& \frac{i}{6} (\dot K_1 K_1^2 + K_1^2 \dot K_1 -
2 K_1 \dot K_1 K_1)     \rangle  \nonumber
\eea

In the case of the uniform tight-binding model (without external potential)  and open boundary conditions the effective Hamiltonians have the following simple form

\be {\cal H}_1 = J^2  D  {\cal Z}_1,  \quad {\cal H}_2 = -\frac{J^3}{3}  (L_3 {\cal U}_1 + L_3^* {\cal B}_1), \ee 
where  $ {\cal Z}_1 =   \delta_{i,j} \delta_{i,1} - \delta_{i,j} \delta_{i,N},$  $N$ is the number of sites,  $D = \sum \limits_{l=1}^{\infty} (|F_l|^2-|F_{-l}|^2)/l$,
${\cal U}_1 = \delta_{i,i+1} \delta_{i,1} +\delta_{i,i+1} \delta_{i,N-1} $ (the upper co-diagonal  with '1' on its ends, and zeros elsewhere),   ${\cal B}_1 =  \delta_{i,i-1} \delta_{i,2} +\delta_{i,i-1} \delta_{i,N}$ (the lower co-diagonal  with '1' on its ends), $L_3$ is a function of driving 
defined in \cite{AINSI}.
In other words, the first-order correction ${\cal H}_1$ is   non-zero only if  perturbation has certain broken time symmetry (for a single harmonic perturbation, $\cos t$, $H_1$ disappears), and  it localizes  near the ends of the chain (near boundaries).  More generally, it  requires non-uniformity of the coupling constant $J$, so  e.g. in a model with alternating coupling constants $J_1 - J_2 - J_1$,  or  with a more general non-uniform coupling $J(n)$, the correction will  be  non-zero throughout  the whole chain \cite{LonghiSI}.
For the uniform coupling $J(n)$, ${\cal H}_2$ is also localized near the boundaries of the open chain.  The  method of multiple time scales \cite{LonghiSI} produces the same results 
in these cases.

A very interesting new result can be obtained for a parabolic external potential $V(n)= V n^2/2$:
the second-order correction to the averaged Hamiltonian corresponds to an induced uniform next-nearest neighbour coupling, which
for the harmonic perturbation with amplitude $K$ is explicitly given by $J' =  - \eps^2 J^2 V \sum \limits_{l=1}^{\infty} \frac{(-1)^l J_l^2(K )}{l^2} $ \cite{AINSI}.

In the case of  infinite uniform lattice, or periodic boundary conditions,  ${\cal H}_1$ and higher-order corrections are absent.

\subsection{Bose-Hubbard model}

We have
\bea
\dot K_1 &=&  - i \{ H  \}  = -i \sum \limits_i \left(  \delta^+ c_i^{\dagger} c_{i+1}  + \delta^- c_{i+1}^{\dagger} c_i     \right), \\
K_1 &=&  -i \sum \limits_i \left(  \delta_1^+ c_i^{\dagger} c_{i+1}  + \delta_1^- c_{i+1}^{\dagger} c_i     \right)
\eea
\be
H K_1 - K_1 H  =   - i  2U \sum \limits_j \Bigl( \delta_1^+ c_j^{\dagger}  (n_j - n_{j+1}) c_{j+1} 
+   \delta_1^-  c_{j+1} ^{\dagger}  (n_{j+1}- n_j )  c_j   \Bigr)
\ee

Since $ [\dot K_1, K_1] = 0$  if we impose periodic boundary conditions,

\bea
i \dot K_2  &=& \{  H K_1 - K_1 H \} ,  \\
K_2 &=&   -2U  \sum \limits_j \left( \delta_2^+ c_j^{\dagger}  (n_j - n_{j+1}) c_{j+1} +  \delta_2^-  c_{j+1} {\dagger}  (n_{j+1}- n_j )  c_j   \right) \nonumber
\eea

For the second-order corrections we have
\bea H K_2 - K_2 H  &=& 2U \sum \limits_{\alpha=1}^6 A_{\alpha},  \nonumber\\
A_1  &=&  - \delta_2^+ \delta_0^+   \sum \limits_j  \Bigl(   c_{j-1}^{\dagger}( 4n_j -n_{j+1} - n_{j-1}) c_{j+1}  -   2 c_j^{\dagger} c_j^{\dagger} c_{j+1} c_{j+1} \Bigr)  = - \delta_2^+ \delta_0^+  a_1 \nonumber\\
A_2 &=& - \delta_2^+ \delta_0^-   \sum \limits_j  \Bigl(  4 n_j n_{j+1} - 2 n_j (n_j-1)  - c_{j-1}^{\dagger} c_{j+1}^{\dagger} c_{j} c_{j}  -c_j^{\dagger} c_j^{\dagger} c_{j+1}   c_{j-1}   \Bigr)  = - \delta_2^+ \delta_0^-  a_2, \nonumber\\
A_3 &=&  \sum \limits_j  \Bigl(4 n_j n_{j+1} - 2 n_j (n_j-1) - c_{j-1}^{\dagger} c_{j+1}^{\dagger} c_{j} c_{j}    - c_j^{\dagger} c_j^{\dagger} c_{j+1}   c_{j-1}         \Bigr) =  - \delta_2^- \delta_0^+  a_3,  \nonumber\\
A_4 &=&  - \delta_2^- \delta_0^-   \sum \limits_j  \Bigl(     c_{j+1}^{\dagger}( 4n_j -n_{j+1} - n_{j-1}) c_{j}  -  2 c_{j+1}^{\dagger} c_{j+1}^{\dagger} c_{j} c_{j} \Bigr)  = - \delta_2^- \delta_0^- a_4,        \\
A_5 &=&  - 2U \delta_2^+  \sum \limits_j  \left( c_j^{\dagger} (n_j - n_{j+1})^2 c_{j+1}  \right)   =  - 2U \delta_2^+  a_5,   \nonumber\\
A_6 &=&  - 2U \delta_2^-  \sum \limits_j  \left( c_{j+1}^{\dagger} (n_j - n_{j+1})^2 c_{j}  \right)  =  - 2U \delta_2^- a_6,    \nonumber
\eea

\bea  -\frac{i}{2} \left( \dot K_1 K_2 - K_2 \dot K_1 \right)  &=& U \sum \limits_{\alpha=1}^4 B_{\alpha},  \nonumber\\
B_1  &=&  \delta^+ \delta_2^+ a_1,   \nonumber \\
B_2  &=&  \delta^+ \delta_2^- a_2, \\
B_3  &=& \delta^- \delta_2^+ a_3,  \nonumber \\
B_4  &=& \delta^- \delta_2^- a_4, \nonumber
\eea

\bea  -\frac{i}{2} \left( \dot K_2 K_1 - K_1 \dot K_2 \right)  &=& U \sum \limits_{\alpha=1}^4 C_{\alpha},  \nonumber\\
C_1  &=&  -( \delta_1^+)^2  a_1,   \nonumber \\
C_2  &=&  - \delta_1^+ \delta_1^- a_2, \\
C_3  &=& - \delta_1^- \delta_1^+ a_3,  \nonumber \\
C_4  &=& -(\delta_1^-)^2  a_4,\nonumber
\eea

\bea  \frac{1}{2}   [[H,K_1],K_1]   &=& U \sum \limits_{\alpha=1}^4 D_{\alpha},  \nonumber\\
D_1  &=&  ( \delta_1^+)^2  a_1,   \nonumber \\
D_2  &=&   \delta_1^+ \delta_1^- a_2, \\
D_3  &=&  \delta_1^- \delta_1^+ a_3,  \nonumber \\
D_4  &=& (\delta_1^-)^2  a_4, \nonumber
\eea

We note that  $D_k = -C_k$ and
time-averages of the coefficients  $A_5$ and $A_6$ are equal to zero.
Also, 
$ [[\dot K_1,K_1],K_1]  = 0.$

We therefore get the effective second-order correction

\bea
{\cal H}_2 &=& U  \sum \limits_{\alpha=1}^4 \Delta_{\alpha} a_{\alpha}, \nonumber\\
\Delta_1 &=& \langle  \delta_2^+  \delta^+ - 2\delta_2^+  \delta_0^+ \rangle,  \nonumber\\
\Delta_2 &=& \langle    \delta^+ \delta_2^-      - 2\delta_2^+ \delta_0^-    \rangle,  \\
\Delta_3 &=& \langle  \delta^- \delta_2^+     - 2\delta_2^- \delta_0^+    \rangle,  \nonumber\\
\Delta_4 &=& \langle   \delta^- \delta_2^-  -   2  \delta_2^- \delta_0^-   \rangle,  \nonumber
\eea

where $a_{\alpha}$ are given by

\bea 
a_1  &=&   \sum \limits_j  \left(   c_{j-1}^{\dagger}( 4n_j -n_{j+1} - n_{j-1}) c_{j+1}  -   2 c_j^{\dagger} c_j^{\dagger} c_{j+1} c_{j+1} \right)  \nonumber\\
a_2 &=&  \sum \limits_j  \Bigl(  4 n_j n_{j+1} - 2 n_j (n_j-1) - c_{j-1}^{\dagger} c_{j+1}^{\dagger} c_{j} c_{j}  
 - c_j^{\dagger} c_j^{\dagger} c_{j+1}   c_{j-1}   \Bigr) \\
a_3 &=&  \sum \limits_j  \Bigl(4 n_j n_{j+1} - 2 n_j (n_j-1) - c_{j-1}^{\dagger} c_{j+1}^{\dagger} c_{j} c_{j}  -c_j^{\dagger} c_j^{\dagger} c_{j+1}   c_{j-1}      \Bigr) \nonumber\\
a_4 &=&    \sum \limits_j  \left(     c_{j+1}^{\dagger}( 4n_j -n_{j+1} - n_{j-1}) c_{j-1}  -   2 c_{j+1}^{\dagger} c_{j+1}^{\dagger} c_{j} c_{j} \right)        \nonumber
\eea

We can simplify it further:

\be {\cal H}_2 = U ( \Delta_1 a_1  + h.c ) +  U(\Delta_2+\Delta_2^*) a_2 \ee

Its not difficult to see that  $(\Delta_2+\Delta_2^*) =- 2 \Delta_0$  as given in the main text,  while $\Delta_1 = -2\Delta^+$.

\subsection{Hubbard model}

We have:
\bea
H  &=&  \sum \limits_{i,\sigma} \Bigl(  \delta_0^+ c_{i,\sigma}^{\dagger} c_{j+1,\sigma}  + \delta_0^- c_{i+1,\sigma}^{\dagger} c_{i,\sigma}     \Bigr)  + U \sum  \limits_{i}  n_{i,\sigma} n_{i,-\sigma}  
\nonumber\\ &=& T^{\uparrow} + T^{\downarrow} + {\cal U} \nonumber\\
\dot K_1 &=& -i \{Ê H \}   =  -i \sum \limits_{j,\sigma} \left(  \delta^+ c_{j,\sigma}^{\dagger} c_{j+1,\sigma}  + \delta^- c_{j+1,\sigma}^{\dagger} c_{j,\sigma}     \right),\nonumber\\
K_1 &=& -i \sum \limits_{j,\sigma} \left(  \delta_1^+ c_{j,\sigma}^{\dagger} c_{j+1,\sigma}  + \delta_1^- c_{j+1,\sigma}^{\dagger} c_{j,\sigma}     \right)  \equiv K_1^{\uparrow} + K_1^{\downarrow} , \nonumber
\eea

\bea
H K_1 - K_1 H  & =&   [T^{\uparrow} , K_1^{\uparrow}  ]  +[T^{\downarrow} , K_1^{\downarrow}  ]   +[  {\cal U}, K_1^{\uparrow}  ] +[ {\cal U}, K_1^{\downarrow}  ] ,\\   T^{\uparrow} K_1^{\uparrow}  -   K_1^{\uparrow} T^{\uparrow} &=&  -i (\delta_0^-  \delta_1^+ - \delta_0^+  \delta_1^- ) \sum \limits_i (n_{i+1,\uparrow} - n_{i,\uparrow}) \nonumber \\
T^{\downarrow}  K_1^{\downarrow} -  K_1^{\downarrow} T^{\downarrow} &=& -i (\delta_0^-  \delta_1^+ - \delta_0^+  \delta_1^- )\sum \limits_i (n_{i+1,\downarrow} - n_{i,\downarrow}) \nonumber \\
{\cal U} K_1^{\uparrow}   -  K_1^{\uparrow}  {\cal U}  &=&  -i U \sum \limits_j (\delta_1^+  c_{j,\uparrow}^{\dagger} c_{j+1,\uparrow}  (n_{j,\downarrow} -n_{j+1,\downarrow} )  -\delta_1^-  c_{j+1,\uparrow}^{\dagger} c_{j,\uparrow}  (n_{j,\downarrow} -n_{j+1,\downarrow} ))  \nonumber\\
{\cal U} K_1^{\downarrow}   -  K_1^{\downarrow}  {\cal U}  &=&  -i U\sum \limits_j (\delta_1^+  c_{j,\downarrow}^{\dagger} c_{j+1,\downarrow}  (n_{j,\uparrow} -n_{j+1,\uparrow} ) -\delta_1^-  c_{j+1,\downarrow}^{\dagger} c_{j,\downarrow}  (n_{j,\uparrow} -n_{j+1,\uparrow} )) \nonumber
\eea

We have also $[\dot K_1, K_1] = 0$, therefore
\bea
i \dot K_2 &=& \{ HK_1-K_1 H  \}, \\
 \dot K_2 &=& \delta_{01} \sum \limits_{j,\sigma} (n_{j,\sigma}-n_{j+1,\sigma}) + U \sum \limits_{j,\sigma} (\delta_1^-  c_{j+1,\sigma}^{\dagger} c_{j,\sigma} -  \delta_1^+ c_{j,\sigma}^{\dagger} c_{j+1,\sigma} )(  n_{j,-\sigma} - n_{j+1,-\sigma}) =
 K_{\delta}^{\uparrow} + K_{\delta}^{\downarrow}+ K_u^{\uparrow} + K_u^{\downarrow} \nonumber
\eea

To calculate  commutator of $H$ and $K_2$, we proceed as follows:
\bea
[H,K_2]  &=& [T^{\uparrow}, K_u^{\uparrow} ] + [ T^{\uparrow}, K_u^{\downarrow}  ]  + [ T^{\downarrow}, K_u^{\uparrow} ]  + [   T^{\downarrow}, K_u^{\downarrow}]  + [ {\cal U}, K_u^{\uparrow}] + [{\cal U}, K_u^{\downarrow}]   \label{HK2} \\
T^{\uparrow}  K_u^{\uparrow}  -  K_u^{\uparrow}  T^{\uparrow}&=&  U \sum \limits_{j}( \delta_0^+ \delta_2^-  + \delta_0^- \delta_2^+ )( 2 n_{j,\uparrow} n_{j,\downarrow} -n_{j,\uparrow} n_{j+1,\downarrow} -n_{j,\downarrow} n_{j+1,\uparrow})  \nonumber\\ 
&+& U\sum \limits_{j}( \delta_0^- \delta_2^-  c_{j+1,\uparrow}^{\dagger}c_{j-1,\uparrow}  + \delta_0^+ \delta_2^+ c_{j-1,\uparrow}^{\dagger} c_{j+1,\uparrow} ) ( n_{j-1,\downarrow} -2n_{j,\downarrow} + n_{j+1,\downarrow})  \nonumber\\
T^{\downarrow}  K_u^{\downarrow}  -  K_u^{\downarrow}  T^{\downarrow}  &=& {\cal P}(\uparrow \to \downarrow,\downarrow \to \uparrow ) ( T^{\uparrow}  K_u^{\uparrow}  -  K_u^{\uparrow}  T^{\uparrow}) \nonumber\\
T^{\uparrow} K_u^{\downarrow} - K_u^{\downarrow}  T^{\uparrow} &=& U \sum \limits_{j} \delta_0^+ \delta_2^- ( - 2 c_{j+1,\downarrow}^{\dagger}  c_{j,\downarrow} c_{j,\uparrow}^{\dagger} c_{j+1,\uparrow}  +  c_{j,\downarrow}^{\dagger}  c_{j-1,\downarrow} c_{j,\uparrow}^{\dagger} c_{j+1,\uparrow} +c_{j+1,\downarrow}^{\dagger}  c_{j,\downarrow} c_{j-1,\uparrow}^{\dagger} c_{j,\uparrow}     )  \nonumber\\ 
&-& U\sum \limits_{j} \delta_0^+ \delta_2^+ ( - 2 c_{j,\downarrow}^{\dagger}  c_{j+1,\downarrow} c_{j,\uparrow}^{\dagger} c_{j+1,\uparrow}  +  c_{j,\downarrow}^{\dagger}  c_{j+1,\downarrow} c_{j-1,\uparrow}^{\dagger} c_{j,\uparrow} +c_{j,\downarrow}^{\dagger}  c_{j+1,\downarrow} c_{j+1,\uparrow}^{\dagger} c_{j+2,\uparrow}    )   \nonumber\\ 
&+& U\sum \limits_{j}\delta_0^-  \delta_2^- (2 c_{j+1,\downarrow}^{\dagger} c_{j,\downarrow} c_{j+1,\uparrow}^{\dagger} c_{j,\uparrow} -c_{j+1,\downarrow}^{\dagger} c_{j,\downarrow} c_{j+2,\uparrow}^{\dagger} c_{j+1,\uparrow} - c_{j+1,\downarrow}^{\dagger} c_{j,\downarrow} c_{j,\uparrow}^{\dagger} c_{j-1,\uparrow}  )  \nonumber\\ 
&-& U\sum \limits_{j} \delta_0^- \delta_2^+  (2 c_{j,\downarrow}^{\dagger} c_{j+1,\downarrow} c_{j+1,\uparrow}^{\dagger} c_{j,\uparrow}  - c_{j,\downarrow}^{\dagger} c_{j+1,\downarrow} c_{j+2,\uparrow}^{\dagger} c_{j+1,\uparrow} - c_{j,\downarrow}^{\dagger} c_{j+1,\downarrow} c_{j,\uparrow}^{\dagger} c_{j-1,\uparrow}  )  \nonumber\\  
T^{\downarrow} K_u^{\uparrow} - K_u^{\uparrow} T^{\downarrow}   &=& U\sum \limits_{j} \delta_0^+ \delta_2^- ( - 2 c_{j+1,\uparrow}^{\dagger}  c_{j,\uparrow} c_{j,\downarrow}^{\dagger} c_{j+1,\downarrow}  +  c_{j+1,\uparrow}^{\dagger}  c_{j,\uparrow} c_{j-1,\downarrow}^{\dagger} c_{j,\downarrow} +c_{j,\uparrow}^{\dagger}  c_{j-1,\uparrow} c_{j,\downarrow}^{\dagger} c_{j+1,\downarrow}    )   \nonumber\\ 
&-& U\sum \limits_{j}\delta_0^+  \delta_2^+ (-2 c_{j,\uparrow}^{\dagger} c_{j+1,\uparrow} c_{j,\downarrow}^{\dagger} c_{j+1,\downarrow}   
        + c_{j,\uparrow}^{\dagger} c_{j+1,\uparrow} c_{j-1,\downarrow}^{\dagger} c_{j,\downarrow}
         + c_{j-1,\uparrow}^{\dagger} c_{j,\uparrow} c_{j,\downarrow}^{\dagger} c_{j+1,\downarrow}  )  \nonumber\\ 
&+& U\sum \limits_{j} \delta_0^- \delta_2^-  (2 c_{j+1,\uparrow}^{\dagger} c_{j,\uparrow} c_{j+1,\downarrow}^{\dagger} c_{j,\downarrow} 
 - c_{j,\uparrow}^{\dagger} c_{j-1,\uparrow} c_{j+1,\downarrow}^{\dagger} c_{j,\downarrow} 
 - c_{j+1,\uparrow}^{\dagger} c_{j,\uparrow} c_{j,\downarrow}^{\dagger} c_{j-1,\downarrow}  )   \nonumber\\
&-& U\sum \limits_{j} \delta_0^- \delta_2^+  (2 c_{j,\uparrow}^{\dagger} c_{j+1,\uparrow} c_{j+1,\downarrow}^{\dagger} c_{j,\downarrow} 
 - c_{j-1,\uparrow}^{\dagger} c_{j,\uparrow} c_{j+1,\downarrow}^{\dagger} c_{j,\downarrow} 
 - c_{j,\uparrow}^{\dagger} c_{j+1,\uparrow} c_{j,\downarrow}^{\dagger} c_{j-1,\downarrow}  )   \nonumber\\
 &=& {\cal P}(\uparrow \to \downarrow,\downarrow \to \uparrow ) ( T^{\uparrow}  K_u^{\downarrow}  -  K_u^{\downarrow}  T^{\uparrow}) \nonumber\\
 {\cal U} K_u^{\uparrow} - K_u^{\uparrow} {\cal U} &=& - U^2 \sum \limits_{j}( \delta_2^-  c_{j+1,\uparrow}^{\dagger} c_{j,\uparrow}   -\delta_2^+  c_{j,\uparrow}^{\dagger} c_{j+1,\uparrow})( n_{j,\downarrow} -n_{j+1,\downarrow} )^2   \nonumber\\
  {\cal U} K_u^{\downarrow} - K_u^{\downarrow} {\cal U} &=& - U^2 \sum \limits_{j}( \delta_2^-  c_{j+1,\downarrow}^{\dagger} c_{j,\downarrow}   -\delta_2^+  c_{j,\downarrow}^{\dagger} c_{j+1,\downarrow})( n_{j,\uparrow} -n_{j+1,\uparrow} )^2   
 \eea
The last two terms in Eq.(\ref{HK2}) will produce zero time-average and are not important for us.
Let us introduce  $[H,K_2]_{\cal U}  \equiv [T^{\uparrow}, K_u^{\uparrow} ] + [ T^{\uparrow}, K_u^{\downarrow}  ]  + [ T^{\downarrow}, K_u^{\uparrow} ]  + [   T^{\downarrow}, K_u^{\downarrow}],$ i.e. the expression  (\ref{HK2}) without the last two terms.
Expressions above represent several fundamental processes.  We introduce the following notation for them:

\bea 
{\cal S} &=&  {\cal S}_{\uparrow, \downarrow}  +  {\cal S}_{  \downarrow, \uparrow} , \quad {\cal S}_{\sigma, -\sigma} =  \sum_j c_{j+1,\sigma}^{\dagger} c_{j,\sigma} c_{j,-\sigma}^{\dagger} c_{j+1,-\sigma},     \\
{\cal A}  &=& {\cal A}_{\uparrow, \downarrow}  +  {\cal A}_{\downarrow, \uparrow },   \quad  {\cal A}_{\sigma, -\sigma} = \sum_j c_{j,\sigma}^{\dagger} c_{j-1,\sigma} c_{j,-\sigma}^{\dagger} c_{j+1,-\sigma}, \nonumber\\
 {\cal R} &=&  {\cal R}_{  \uparrow \downarrow} +  {\cal R}_{ \downarrow \uparrow},  \quad {\cal R}_{ \sigma,- \sigma} = \sum_j c_{j,\sigma}^{\dagger} c_{j-1,\sigma} c_{j+1,-\sigma}^{\dagger} c_{j,-\sigma},  \nonumber\\
 {\cal \bf R} &=&  \sum_j c_{j+1,\downarrow}^{\dagger} c_{j,\downarrow} c_{j+1,\uparrow}^{\dagger} c_{j,\uparrow} ,  \nonumber\\
  R_2 &=&  R_{2,\uparrow} + R_{2,\downarrow}, \quad R_{2,\sigma} =  \sum_j c_{j+1,\sigma}^{\dagger} c_{j-1,\sigma}  (n_{j-1,-\sigma}- 2n_{j,-\sigma} +n_{j+1,-\sigma}),   \nonumber\\
V & =& \sum_{\sigma} V_{\sigma},  \quad  V_{\sigma} = \sum_j n_{j,\sigma} n_{j+1,-\sigma}, \nonumber\\
 {\cal \bf V} &=& \sum_j n_{j,\downarrow} n_{j,\uparrow}
 \eea
 There are several different types of tunnelling and interaction processes here (${\cal S, A,R}, {\bf R}, R_2, V,$ and ${\cal \bf V}$ ). ${\cal S}$ denotes a correllated tunneling process of two atoms on neighboring sites exchanging their positions ('superexchange'). ${\cal A}^{\dagger}$   denotes  a  tunnelling process where a pair of atoms at $j$th site is dissociated into two atoms at neighboring sites ($j-1$ and $j+1$).
  ${\cal A}$   denotes a corresponding correlated tunneling process of association of two atoms  into a pair.
  ${\cal R}$   denotes  a correlated tunnelling process where an extended pair of atoms is tunnelling to the right neighboring sites,
  and ${\cal R}^{\dagger}$ denotes tunnelling to the left. $ {\cal \bf R}$  denotes tunnelling of a localized pair of atoms to the right, and   ${\cal \bf R}^{\dagger}$ denotes tunnelling to the left. $R_2$ denotes  single-particle next-nearest-neighbour density-dependent tunnelling to the right, and  $R_2^{\dagger}$ denotes corresponding tunelling to the left.
  $V$ denotes nearest-neighbour interaction, and  $ {\cal \bf V}$ denotes usual local interaction.
  We can rewrite $[H,K_2]_{\cal U}$ in terms of these fundamental processes:
  
 \bea
 [H,K_2]_{\cal U} = U [   
 \left(\delta_0^- \delta_2^- (4  {\cal \bf R} -2{\cal R}  +R_2 ) + H.c. \right) +  (\delta_0^+ \delta_2^-   +  \delta_0^- \delta_2^+ ) (-2{\cal S} + {\cal A} +{\cal A}^{\dagger}+4 {\cal \bf V} -2 V ) ]
 \eea 
 
 Now, let us consider the term $\frac{1}{2}[[H,K_1],K_1]$.
 It can be shown  $\frac{1}{2}[[H,K_1],K_1] = \frac{1}{2}{\cal P}(\delta_0^{\pm} \to \delta_1^{\pm}, \delta_2^{\pm} \to -\delta_1^{\pm}) [H,K_2]_{\cal U}$.
 Also, \be-\frac{i}{2}[\dot K_2,K_1]  =\frac{1}{2}{\cal P}(\delta_0^{\pm} \to \delta_1^{\pm}, \delta_2^{\pm} \to \delta_1^{\pm}) [H,K_2]_{\cal U}= -\frac{1}{2}[[H,K_1],K_1], \ee
   and
   \be
   -\frac{i}{2}[\dot K_1,K_2] =-\frac{1}{2} {\cal P}(\delta_0^{\pm} \to \delta^{\pm}) [H,K_2]_{\cal U}
   \ee 
   
   We have finally
   
\bea
{\cal H}_2 &=& U  \Delta_0 (-2{\cal S} + {\cal A} +{\cal A}^{\dagger} + 4 {\cal \bf V} -2 V )+ U \Bigl[   \Delta^-  (4  {\cal \bf R} -2{\cal R}  +R_2 ) +H.c. \Bigr],\nonumber
\eea
where
\bea
 \Delta_0   &=& \langle  \delta_2^- \left(\delta_0^+ - \frac{1}{2} \delta^+ \right)  + \delta_2^+ \left(\delta_0^- - \frac{1}{2}\delta^- \right)   \rangle = \frac{1}{2}  \langle  \delta_2^- \delta^+ + \delta^-  \delta_2^+  \rangle \\
  \Delta^-  &=&  \langle  \delta_2^- \left(\delta_0^- - \frac{1}{2} \delta^- \right)    \rangle  =  \frac{1}{2} \langle  \delta_2^- \delta^-   \rangle   = (\Delta^+ )^*  \nonumber
  \eea
\subsection{Driven DNLSE as a classical limit of driven Bose-Hubbard model }
  
  Higher-order corrections for the driven Bose-Hubbard model  discussed above   
  can be compared with the corrections for a driven DNLSE obtained using canonical perturbation theory. 
  In the limit of large occupation numbers one would expect to obtain similar results for both systems.  
  Indeed,  consider a fastly driven DNLSE
 
\be
i \dot \psi_n = -J( \psi_{n-1} + \psi_{n+1})  + g |\psi_n|^2 \psi_n  -  \omega E(\omega t) n \psi_n
\ee

We remove the last term using the transformation $ \psi_n =v_n \exp[-i n \omega \int \limits_0^t E(\omega t') dt']$,  obtaining
in new variables

\be
i \dot v_n = - J (v_{n+1} F(\omega t) + v_{n-1} F^*(\omega t))  + g |v_n|^2 v_n , 
\ee
 $ F(\omega t) =  \exp [-i \omega \int \limits_0^t E(\omega t') dt']$
This is an equation corresponding  to a classical Hamiltonian

\be
H= - J \sum \limits_n [ F v_n^* v_{n+1} + F^* v_n^* v_{n-1} ] + \frac{g}{2} \sum \limits_n |v_n|^4,
\ee
introducing the fast time $ \tau = \omega t \equiv t/\eps$, we  get the Hamiltonian $\eps H$,
where $F(\tau) =  \exp [-i \omega \int \limits_0^t E(\omega t') dt']  = \exp [-i \int \limits_0^{\tau} E(\tau) d \tau]  =  
F_0 + \sum F_l \exp (-i l \tau)$.   Obviously,  $ |v_n|^2 = |\psi_n|^2$, so the physical meaning of new and old variables is very similar.

Now,  we apply a canonical transformation using the generating function

\be
W =  i \sum\limits_n v_n V_n^*  +  i \eps S({\bf v, V^*}, \tau),  \quad S =   S_1 ({\bf v, V^*},\tau)  + \eps S_2 ({\bf v,V^*},\tau) +..,
\ee
where $\{\bf v,V^* \}$ denote the whole set of variables $v_n,V_n^*, \quad n=1,..N$

The new and  the old variables are related by 

\bea
v_n^* = -i \frac{\partial W}{v_n} =  V_n^*  + \eps \frac{ \partial S_1}{\partial v_n} + 
\eps^2 \frac{\partial S_2}{\partial v_n}  \nonumber\\
V_n = -i \frac{ \partial W}{\partial V_n^*}  = v_n + \eps \frac{\partial S_1}{\partial V_n^*} + \eps^2 \frac{\partial S_2}{\partial V_n^*}
\eea
We want to obtain an effective time-independent Hamiltonian $\eps {\cal H} $ using  this transformation.
The effective Hamiltonian can be expanded in series in $\eps$  

\be
{\cal H}({\bf V,V^* }) =  {\cal H}_0({\bf V, V^*}) + \eps {\cal H}_1({\bf V,V^* }) + ..  
\ee

After tedious calculations  determing appropriate generating function, one obtains \cite{AINPrSI}

\bea
{\cal H}_1 &=& 0, \nonumber\\
{\cal H}_2 &=&  -J^2 g {\Phi} \left(  v_n^2 v_{n-1}^* v_{n+1}^* +  ( v_n^* )^2 v_{n+1} v_{n-1} +2 |v_n|^4 - 4|v_n|^2 |v_{n+1}|^2 \right) \nonumber\\ &+&  J^2 g T \left(  (|v_{n-1}|^2 - 2|v_n|^2) v_{n+1} v_{n-1}^* + v_{n+1}^2 (v_n^*)^2   \right) \nonumber\\
&+&  J^2 g T^* \left(  (|v_{n+1}|^2 - 2|v_n|^2) v_{n+1}^* v_{n-1} + v_{n}^2 (v_{n+1}^*)^2   \right)  \nonumber\\
&+& J^3 (.....),
\eea
where $\Phi \equiv  \langle |\tilde F|^2 \rangle = \sum \limits_{l \ne 0} \frac{|F_l|^2}{l^2}$,  $\tilde F = i \sum \limits_{l \ne 0} \frac{F_l}{l} \exp(-il t')$,
$T  \equiv -\sum \limits_{l\ne0} \frac{F_l F_{-l}}{l^2}$.

If we now consider expressions obtained earlier for Bose-Hubbard models,  and assume Fourier expansion $\delta_0^+ = F_0 + \sum \limits_{l \ne 0} F_l \exp (-il t)$,
we obtain exactly $\Delta_0 = - J^2 \Phi$,  $\Delta^+ = \frac{1}{2} J^2 T$.  Therefore, the semiclassical limit of our averaging procedure for the Bose-Hubbard model  produces the result analogous  to the canonical perturbation theory for DNLSE (where variables $v_n$ are considered as c-number analogs of the creation/annihilation operators of the Bose-Hubbard model).
 
\begin{figure*}[h]
\includegraphics[width=80mm]{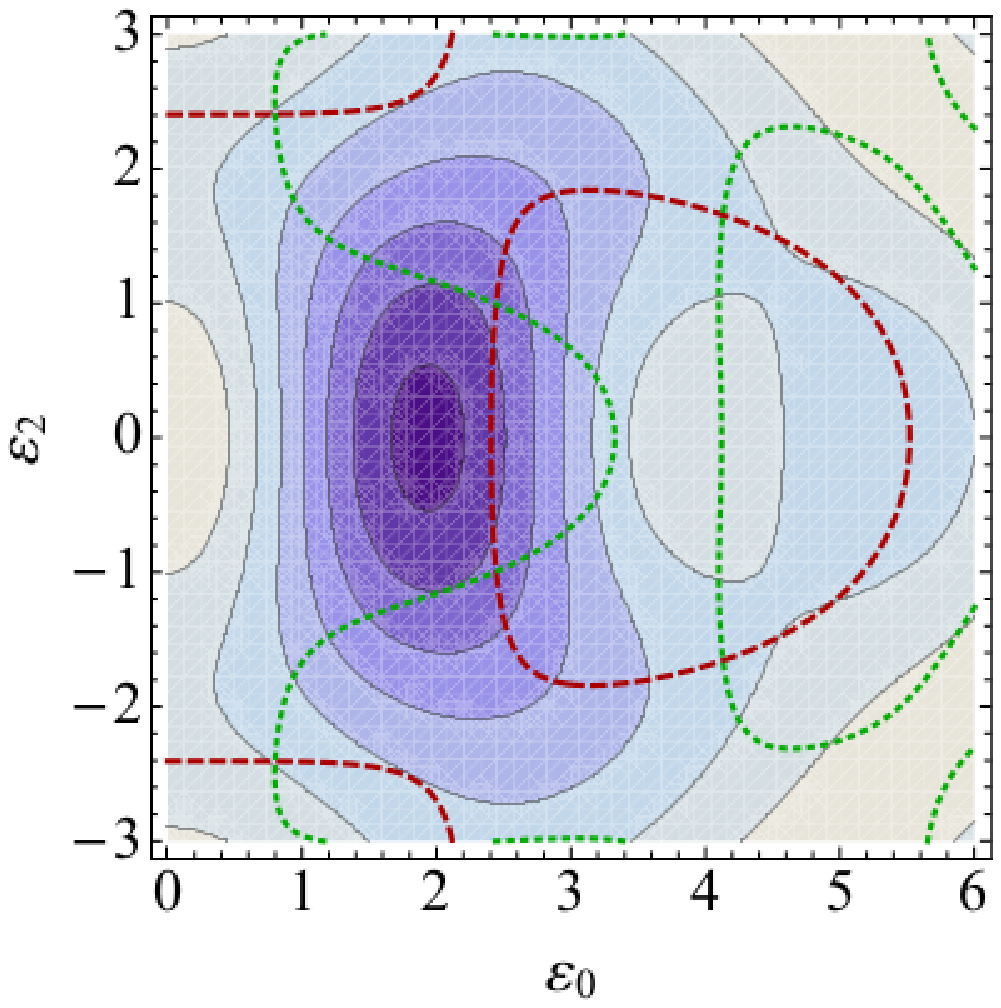}
\includegraphics[width=80mm]{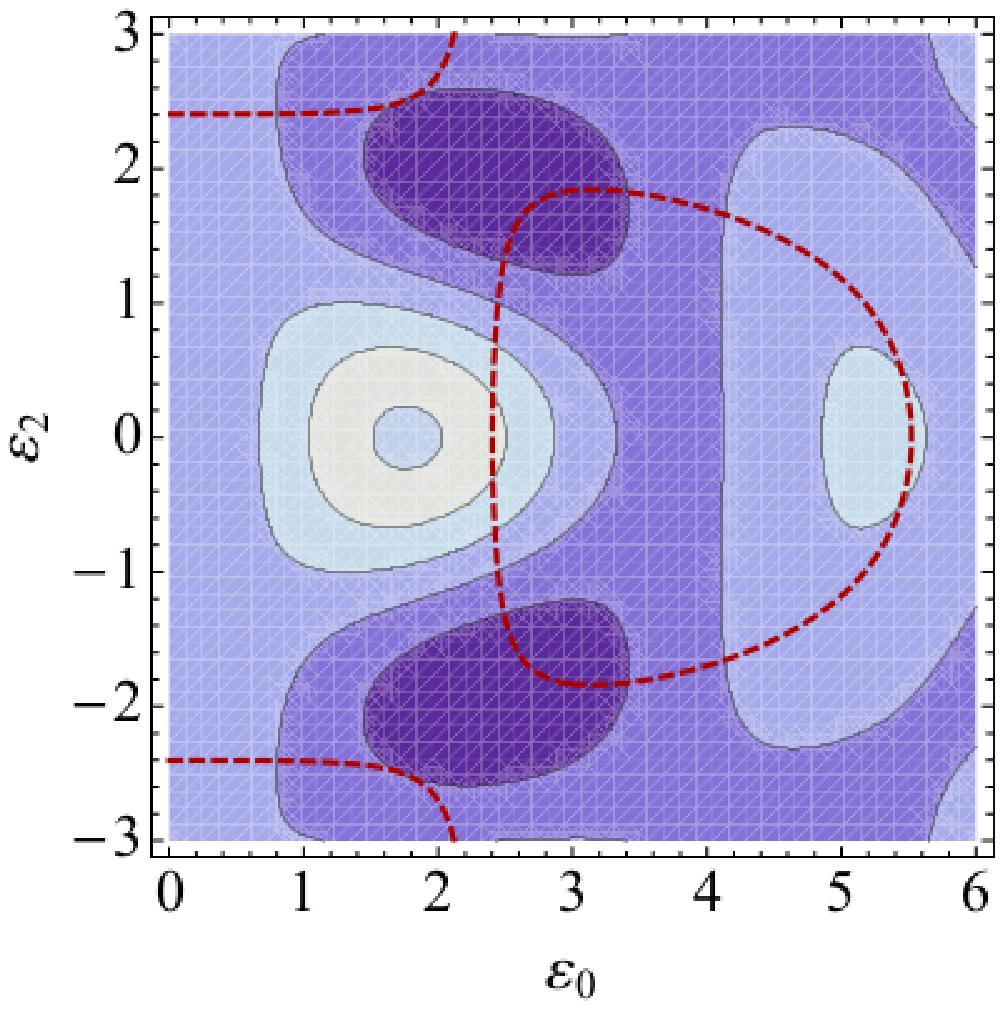}
\includegraphics[width=80mm]{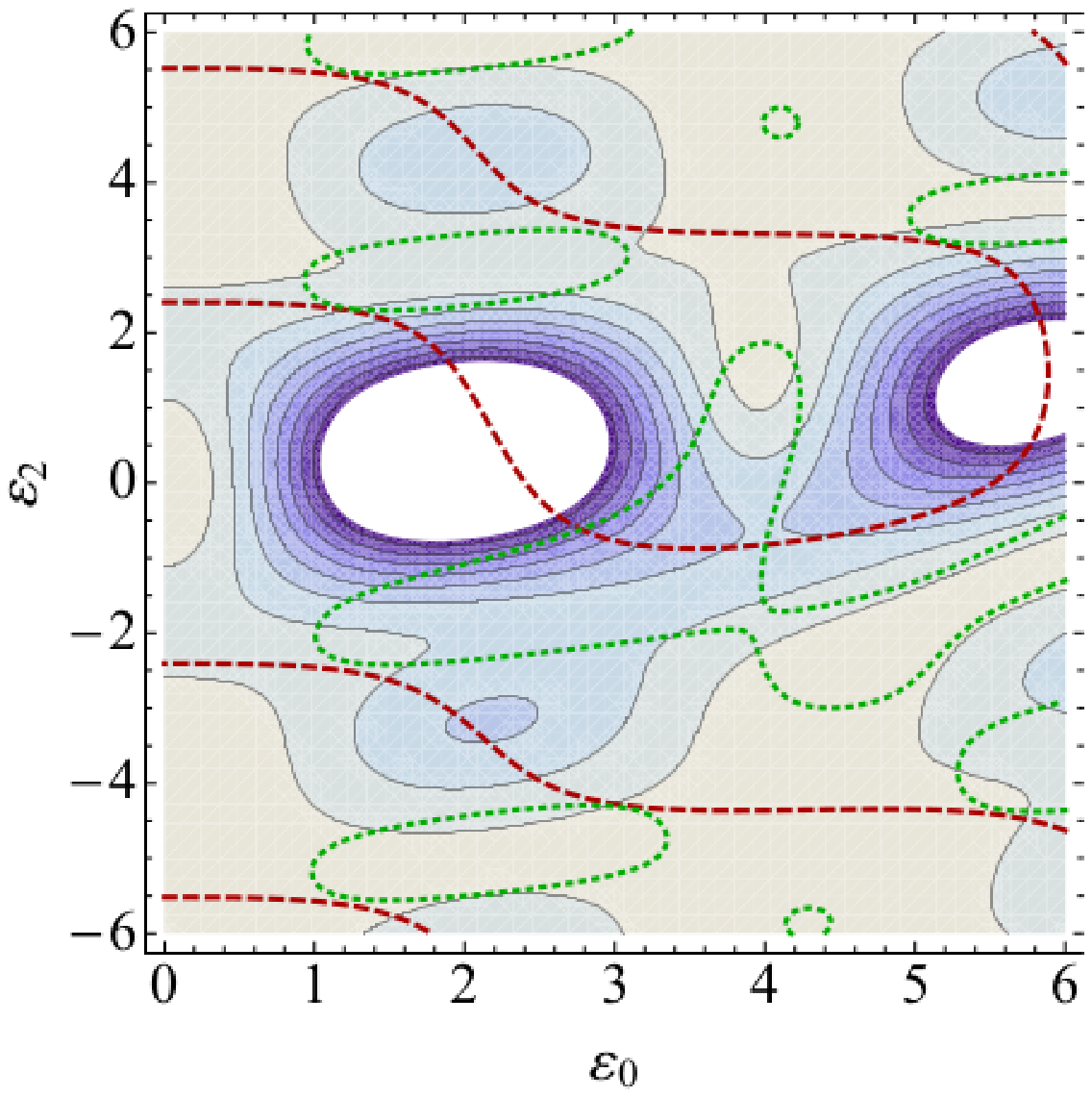}
\includegraphics[width=80mm]{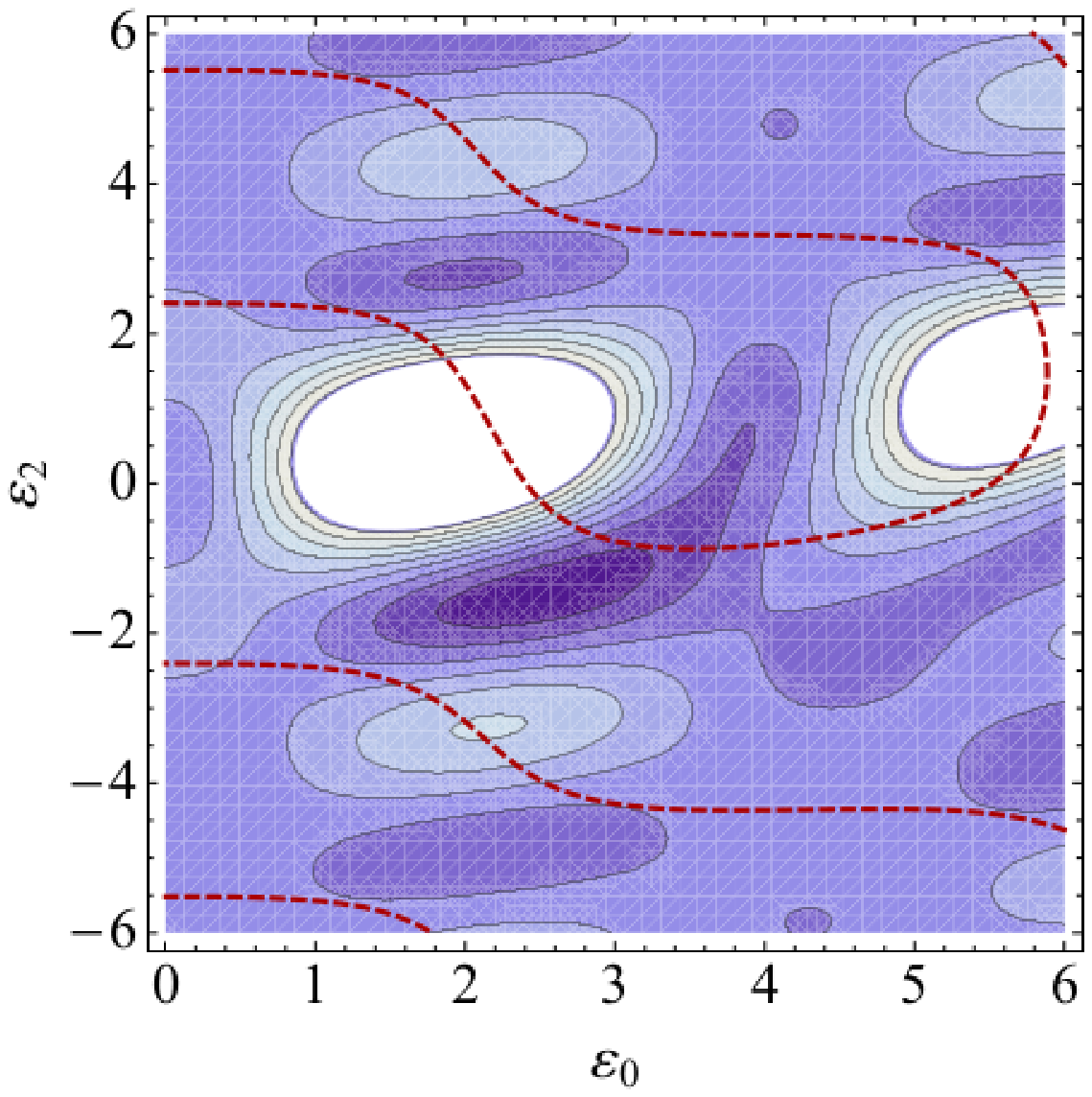}
\caption{ (Color online)
Parameters  $\Delta_0$ (left panel), $\Delta$ (right panel) as a function of  ${\cal E}_0,{\cal E}_2$.Upper panel: two-colour driving with double frequency  $f(t) =  {\cal E}_0 \sin t + {\cal E}_2 \sin 2t$,
bottom panel:  two-colour driving with tripled frequency  $f(t) =  {\cal E}_0 \sin t + {\cal E}_2 \sin 3t$. Dotted (green) line:  zeros of $\Delta = \Delta^+ + \Delta^-$  which controls pair tunnelling.   Dashed (red) line:  zeros of effective single-particle tunnelling $J_e$.
 \label{Fig25SI}}
\end{figure*}
\subsection{Engineering the parameters of the effective Hamiltonian.}

Here we show how a simple non-monochromatic perturbation allows to vary parameters $\Delta_0$, $\Delta^{\pm}$ of the effective Hamiltonian.

Let $f(t) =  {\cal E}_0 \sin t + {\cal E}_2 \sin 2t$,  i.e. a two-parameter, bichromatic driving. 
Then,
\bea \delta_0^+ &=&  J \exp \Bigl[ i ({\cal E}_0 \sin t + {\cal E}_2 \sin 2t )  \Bigr]  = J  \sum \limits_n J_n ({\cal E}_0) e^{i nt}  \sum \limits_m J_m ({\cal E}_2) e^{i 2 mt}, \nonumber\\
\langle \delta_0^+ \rangle &=& J   \sum \limits_m J_m ({\cal E}_2) J_{-2m}({\cal E}_0), \quad  \delta^+ =  J  \sum \limits_{n \ne -2m} \sum \limits_m J_n ({\cal E}_0)J_m ({\cal E}_2)  e^{i (n+2m)t} ,   \\
\langle \delta_2^+ \delta^-  \rangle  &=& -J^2 \sum_{n_1 \ne -2m_1} \sum\limits_{m_1,m_2} \frac{ J_{n_1}  ({\cal E}_0)  J_{m_1}({\cal E}_2) J_{m_2}({\cal E}_2)  J_{n_1+2m_1-2m_2}({\cal E}_0) }{(n_1+2m_1)^2} 
\eea 
 
 Plotting  $\Delta_0,  \Delta= \Delta^- + \Delta^+$ as a function of  parameters ${\cal E}_0,{\cal E}_2$ (Fig 2-5),
 one  notes that it is possible to  (i)  nullify  single-particle and doublon tunnelling  simultaneously, while keeping exchange interaction  large (ii) nullify single-particle tunnelling keeping
 only two-particle tunnelling and exchange interactions (iii)  nullify  doublon tunnelling while keeping single-particle tunnelling finite.

 \subsection{Driven fermionic two-site model}
 
 For the fermonic  two-site model  we wish to obtain even higher order (fourth order and beyond) corrections.
Expressions for them become very bulky.
However, in the case of strong interaction (large $U$),  we can keep only the terms of the highest order in $U$ in all terms of fourth order in $1/\omega$ and higher.
Such  diagrammatic-like technique allows to obtain very accurate expressions for the effective Hamiltonian. 
In the expressions below the subscript  $_U$  denotes the leading term in $U$.
So,  starting with 

\be
H = \delta_0^+(t) ( c_{1\uparrow}^{\dagger}  c_{2 \uparrow} + c_{1\downarrow}^{\dagger}  c_{2 \downarrow}) +
\delta_0^-(t) ( c_{2\uparrow}^{\dagger} c_{1\uparrow}  +  c_{2\downarrow}^{\dagger} c_{1\downarrow} ) + U (n_{1\uparrow} n_{1 \downarrow}  +n_{2\uparrow } n_{2\downarrow} )
\ee

we have

\be
{\cal H }_0= \langle H \rangle = J_0 ( c_{1\uparrow}^{\dagger}  c_{2 \uparrow} + c_{1\downarrow}^{\dagger}  c_{2 \downarrow}) +
J_0 ( c_{2\uparrow}^{\dagger} c_{1\uparrow}  +  c_{2\downarrow}^{\dagger} c_{1\downarrow} ) + U (n_{1\uparrow} n_{1 \downarrow}  +n_{2\uparrow } n_{2\downarrow} )
\approx  U (n_{1\uparrow} n_{1 \downarrow}  +n_{2\uparrow } n_{2\downarrow} ) \equiv  \langle H \rangle_U
\ee

\be
i \dot K_1 = \{ H \}, \quad
K_1 = -i \left(  \delta_1^+ ( c_{1\uparrow}^{\dagger} c_{2\uparrow} + c_{1\downarrow}^{\dagger}  c_{2 \downarrow}  )  +  \delta_1^-  (c_{2\uparrow}^{\dagger} c_{1\uparrow}  +  c_{2\downarrow}^{\dagger} c_{1\downarrow}  )  \right) \sim  U^0
\ee

\be
[ \{ H \}, K_1 ] = (\delta_1^+ \delta^- - \delta_1^- \delta^+) ] (n_1-n_2)
\ee

\bea
i \dot K_2 &=& \{ [H -\frac{\{H \}}{2},K_1] \}  =  \{ [\langle H \rangle +\frac{\{H \}}{2},K_1] \}  \nonumber\\ & =&  [ \langle H \rangle+\frac{\{H \}}{2},K_1]    -
 \frac{1}{2} \langle  [ \{ H \} , K_1]   \rangle  =  [ \langle H \rangle+\frac{\{H \}}{2},K_1] \approx   [ \langle H \rangle_U,K_1] 
\eea

\be
{\cal H}_1 = \frac{1}{2} \langle  [ \{ H\}, K_1]   \rangle  = 0
\ee

\bea
K_2  &=&  - \delta_{02} ( n_{2\uparrow} - n_{1\uparrow} + n_{2\downarrow} - n_{1\downarrow})  + U \sum\limits_{\sigma} (n_{1 ,\sigma}-n_{2,-\sigma})(\delta_2^-
c_{2,\sigma}^{\dagger} c_{1,\sigma} - \delta_2^+ c_{1,\sigma}^+ c_{2,\sigma})  \nonumber\\ &\approx&
U \sum\limits_{\sigma} (n_{1 ,\sigma}-n_{2,-\sigma})(\delta_2^-
c_{2,\sigma}^{\dagger} c_{1,\sigma} - \delta_2^+ c_{1,\sigma}^+ c_{2,\sigma})  \equiv K_{2U}
  \eea

\be
{\cal H}_2 = \frac{1}{2} \langle  [\{ H \}, K_2]  \rangle +\frac{1}{12} \langle [ \{ [\{ H\},K_1] \} ,K_1]  \rangle   \label{H2}
\ee

Since
\be
\langle [ \{ [\{ H\},K_1] \} ,K_1]  \rangle = 2i \langle \delta_1^- (\delta_1^+ \delta^- - \delta_1^- \delta^+) \rangle ( c_{2\downarrow}^{\dagger} c_{1\downarrow} + c_{2\uparrow}^{\dagger}c_{1\uparrow}) + h.c,
\ee
the last term in Eq.(\ref{H2}) adds a small correction to effective tunnelling constant $J_{eff}=J_0$. This is a finite size effect which is absent in infinite 1D lattice or
1D lattice with periodic boundary conditions.
The first term gives

\bea
\frac{1}{2} \langle  [\{ H \}, K_2]  \rangle &=&  \Bigl[ \langle \delta_{02} \delta^- \rangle (c_{2\downarrow}^{\dagger} c_{1\downarrow} + c_{2\uparrow} c_{1\uparrow}) +h.c \Bigr] \nonumber\\ &+&
 \langle  (\delta_2^+ \delta^- + \delta_2^- \delta^+) \rangle U   \Bigl[ (n_{1\downarrow} - n_{2\downarrow})(n_{1\uparrow}-n_{2\uparrow}) - (c_{2\uparrow}^{\dagger}  c_{1\uparrow} c_{1\downarrow}^{\dagger} c_{2\downarrow} +  c_{2\downarrow}^{\dagger}  c_{1\downarrow} c_{1\uparrow}^{\dagger} c_{2\uparrow} ) \Bigr]  \nonumber\\
 &- & 2   \Bigl[ \langle \delta_2^- \delta^- \rangle U  d_2^{\dagger} d_1  +h.c. \Bigr],
\eea
where we introduce doublon operators $d_k = c_{k\downarrow} c_{k\uparrow}$, $k=1,2$.

While general expressions for $ {\cal H}_k, K_k$ become very complicated with increasing $k$, leading term in $U$ can be calculated:

\be
{\cal H}_3  \approx  \langle \delta_1^+ \delta_2^-  -  \delta_1^- \delta_2^+  \rangle ( n_{1\downarrow} n_{1\uparrow} -..) = 0,  \quad \mbox{if} \quad  |F_l|^2 = |F_{-l}|^2
\ee

\bea
{\cal H}_4 & \approx &  - U^3 \langle  \delta_4^- \delta^+ + \delta_4^+ \delta^-  \rangle \Bigl[ (n_{1\downarrow} - n_{2\downarrow})(n_{1\uparrow}-n_{2\uparrow}) - (c_{2\uparrow}^{\dagger}  c_{1\uparrow} c_{1\downarrow}^{\dagger} c_{2\downarrow} +  c_{2\downarrow}^{\dagger}  c_{1\downarrow} c_{1\uparrow}^{\dagger} c_{2\uparrow} ) \Bigr] \nonumber\\ 
&+& 2U^3 \Bigl[ \langle   \delta_4^- \delta^-  \rangle    d_2^{\dagger} d_1  +h.c. \Bigr]
\eea
 
I.e.,  the leading term in ${\cal H}_4$ has exactly the same structure as in ${\cal H}_2$,  just with different coefficients.  It is not difficult to sum up  infinite number of leading terms of ${\cal H}_k$ with the same structure.

For the harmonic perturbation  used in \cite{MentinkSI}  one has  

\bea
\langle \delta_4^+ \delta^- \rangle &=& \langle \delta_4^- \delta^+ \rangle = \frac{A_4}{2}  = 2 J^2 \sum \limits_{m=1}^{\infty} \frac{J_m^2({\cal E}_0)}{m^4}  \nonumber\\
\langle \delta_4^- \delta^- \rangle &=& \langle \delta_4^+ \delta^+ \rangle = \frac{C_4}{2}  = 2 J^2 \sum \limits_{m=1}^{\infty} \frac{ (-1)^m J_m^2({\cal E}_0)}{m^4} \label{deltas}\\
\langle  \delta_2^+ \delta^- \rangle &=& \langle  \delta_2^- \delta^+ \rangle = \frac{A_2}{2}  = -2 J^2 \sum \limits_{m=1}^{\infty} \frac{J_m^2({\cal E}_0)}{m^2} \nonumber\\
\langle  \delta_2^- \delta^- \rangle &=& \langle  \delta_2^+ \delta^+ \rangle = \frac{C_2}{2}  = -2 J^2 \sum \limits_{m=1}^{\infty} \frac{(-1)^m J_m^2({\cal E}_0)}{m^2} \nonumber
\eea

\bea
{\cal H}_{eff}  &\approx &  {\cal H}_0  +  \left( A_2 \frac{U}{\omega^2}  - A_4 \frac{U^3}{\omega^4} +..  \right) \Bigr[  (n_{1\downarrow} - n_{2\downarrow})(n_{1\uparrow}-n_{2\uparrow}) - (c_{2\uparrow}^{\dagger}  c_{1\uparrow} c_{1\downarrow}^{\dagger} c_{2\downarrow} +  c_{2\downarrow}^{\dagger}  c_{1\downarrow} c_{1\uparrow}^{\dagger} c_{2\uparrow} )  \Bigl] \nonumber\\  &-&  \left( C_2 \frac{U}{\omega^2}  - C_4 \frac{U^3}{\omega^4} +..  \right) ( d_2^{\dagger} d_1 + h.c.)
\eea

Denoting expressions in  round brackets as $A$ and $C$, correspondingly,  we get the following  matrix elements of the effective Hamiltonian in the basis of 4 states 
of the two-site system:

\be
{\cal H}_{eff}  =   \left( \begin{array}{cccc}
-A & -A & J_{e} &J_{e} \\
-A & -A & J_{e}  & J_{e}\\
J_{e} & J_{e} &  U+A  & -C \\
J_{e} & J_{e} &  -C  & U+A   \end{array} \right),
\ee
where $J_e$ denotes effective tunnelling.

The half-difference between the two lowest levels is
\be
\frac{\Delta E}{2} =  J_{ex} = \frac{1}{4} \left(  A+C-U + \sqrt{ 16J_e^2 + (U+3A -C)^2 }   \right) \approx  \frac{2J_e^2}{U} + A + \frac{2J_e^2}{U^2} (C-3A) + ..  \label{theory}
\ee

For the harmonic perturbation summation of infinite number of  leading order in $U$ terms  leads to

\be
A  = -4J^2 \sum \limits_{m=1}^{\infty} J_m^2({\cal E}_0) \left(  \frac{U}{\omega^2 m^2} + \frac{U^3}{\omega^4 m^4} +.. \right) = - 4J^2 U \sum \limits_{m=1}^{\infty} \frac{J_m^2({\cal E}_0)}{m^2 \omega^2 - U^2}   \label{infA}
\ee

Note that although expression $(\ref{infA})$ can change its sign at  $\omega < U$, it was obtained as a summation of a geometric series  with a multiplicator  $\frac{U^2}{m^2 \omega^2}$,
therefore it is assumed  $ \omega > U$.    The case   $\omega < U$ deserves further consideration,  in particular because resonances between $m \omega$ and $U$ can happen.
It is a remarkable fact however that  this formal expression  is correct also  for $\omega < U$, according to \cite{MentinkSI}
 
\begin{figure}
\includegraphics[width=140mm]{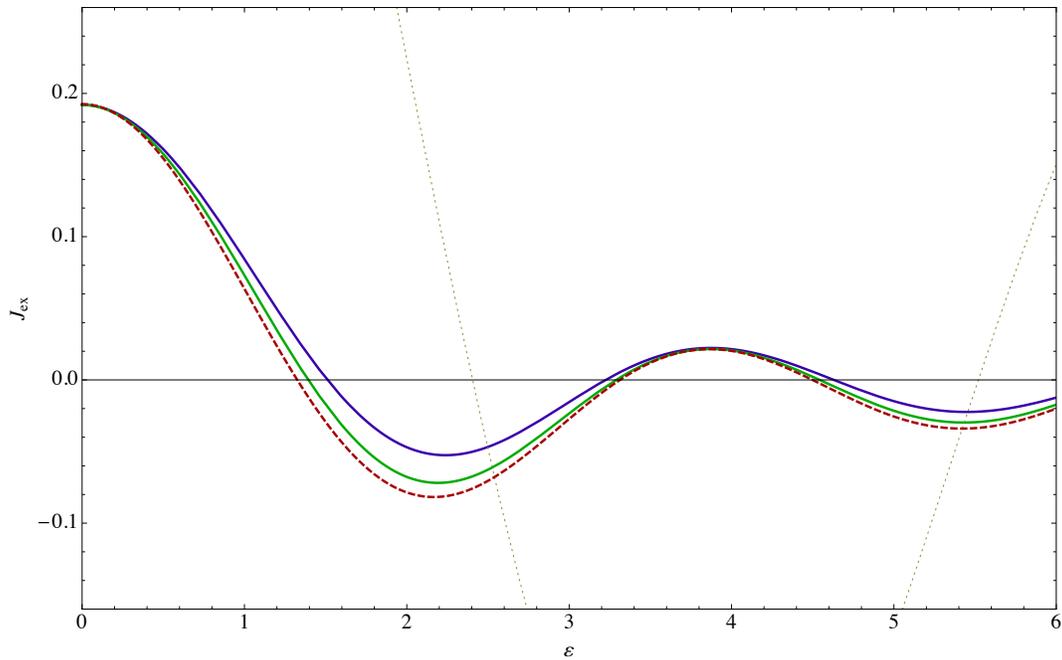}
%\epsfbox{FF1.eps}
\caption{ (Color online)
Induced exchanged interaction in the driven fermionic two-site model. Solid curves,  from up to down:  theoretical prediction Eq. (\ref{theory}) with corrections up to the second order in $\frac{1}{\omega}$  taken into account (upper curve);  with corrections up to the fourth order in $\frac{1}{\omega}$  taken into account (lower curve). Dashed curve:  numerical value of the exchange interaction (numerical data kindly provided by J.Mentink).  Dotted curve:  effective tunnelling  $J_e$.  It is seen that in the regions  where  $J_e$  is strongly suppressed due to driving,  second order corrections become insufficient, and one need to include fourth and higher orders in $\frac{1}{\omega}$.
 \label{Fig1SI} }
\end{figure}

%Naturally, it is  interesting to know if it is possible to enhance control of exchange interactions by using  a bichromatic or a more complicated driving.

%%  Insert SI

\begin{thebibliography}{10}
\bibitem{Stephenson2}
Stabilization of unstable equilibrium of a pendulum with high-frequency driving was studied firstly in \cite{Stephenson} using a linearization procedure, i.e. without introducing effective potentials (being elaborated in Refs. \cite{Bogolyubov,Kapitza}).
\end{thebibliography}
\end{document}